\documentclass{aa}  
\usepackage{graphicx}
\usepackage{txfonts}
\begin{document}
\title{GMASS Ultradeep Spectroscopy of Galaxies at $1.4<z<2$. II. Superdense passive galaxies: how did they form and evolve ?\thanks{Based on the ESO VLT Large Program 173.A--0687}}

\author{A. Cimatti
\inst{1} 
\and
P. Cassata\inst{2} \and	
L. Pozzetti\inst{3} \and
J. Kurk\inst{4} \and
M. Mignoli\inst{3} \and
A. Renzini\inst{5} \and
E. Daddi\inst{6} \and
M. Bolzonella\inst{3} \and
M. Brusa\inst{7} \and
G. Rodighiero\inst{8} \and
M. Dickinson\inst{9} \and
A. Franceschini\inst{8} \and
G. Zamorani\inst{3} \and
S. Berta\inst{6} \and
P. Rosati\inst{10}
C. Halliday\inst{11}
}
	   
\offprints{A. Cimatti}

\institute{Dipartimento di Astronomia, Universit\`a di Bologna, 
Via Ranzani 1, I-40127, Bologna, Italy.
\email{a.cimatti@unibo.it}
\and   
LAM, Traverse du siphon, 13012 Marseille, France
\and		
INAF- Osservatorio Astronomico di Bologna,
Via Ranzani 1, I-40127, Bologna, Italy
\and		
Max-Planck-Institut f\"ur Astronomie, K\"onigstuhl 17, 
D-69117, Heidelberg, Germany
\and
INAF- Osservatorio Astronomico di Padova,
Vicolo dell'Osservatorio 5, I-35122, Padova, Italy
\and
CEA-Saclay, DSM/DAPNIA/Service d'Astrophysique, 91191 
Gif-sur-Yvette Cedex, France
\and
Max Planck Institut f\"ur Extraterrestrische Physik, 
Postfach 1312, 85741 Garching bei M\"unchen, Germany
\and
Universit\`a di Padova, Dipartimento di Astronomia, 
Vicolo dell'Osservatorio 2, I-35122, Padova, Italy
\and
NOAO-Tucson, 950 North Cherry Avenue, Tucson, AZ 85719, USA
\and
European Southern Observatory, 
Karl-Schwarzschild-Strasse 2, D-85748, Garching bei M\"unchen, Germany
\and
INAF- Osservatorio Astrofisico di Arcetri,
Largo E. Fermi 5, I-50125, Firenze, Italy 
}

\date{Received ....., 2007; accepted ........, 2007}

\abstract
{}
{The aim of this work is to investigate the physical, structural and 
evolutionary properties of old, passive galaxies at $z>1.4$ and to
place new constraints on massive galaxy formation and evolution.}
{We combine ultradeep optical spectroscopy from the GMASS project 
(\emph{Galaxy Mass Assembly ultradeep Spectroscopic Survey}) with 
GOODS multi-band (optical to mid--infrared) photometry and HST imaging
to study a sample of spectroscopically identified passive galaxies
at $1.39<z<1.99$ selected from \emph{Spitzer Space Selescope} imaging
at 4.5$\mu$m.}
{A stacked spectrum with an equivalent integration time of $\sim$
500 hours was obtained and compared with libraries of synthetic stellar
population spectra. The stacked spectrum is publicly released.
The spectral and photometric SED properties indicate very weak or absent 
star formation, moderately old stellar ages of $\approx$1 Gyr 
(for solar metallicity) and stellar masses in the range of $10^{10-11}$ 
M$_{\odot}$, thus implying that the major star formation and assembly 
processes for these galaxies occurred at $z>2$. No X-ray emission was 
found neither from individual galaxies nor from a stacking analysis 
of the sample. Only one galaxy shows a marginal detection at 24$\mu$m. 
These galaxies have morphologies that are predominantly compact and spheroidal.
However, their sizes ($R_e \lesssim$ 1 kpc) are much smaller than those
of spheroids in the present--day Universe. Their stellar mass surface 
densities are consequently higher by $\approx$1 dex if compared to 
spheroids at $z\approx0$ with the same mass. Their rest-frame $B$-band
surface brightness scales with the effective radius, but the offset 
with respect to the surface brightness of the local Kormendy relation is 
too large to be explained by simple passive evolution. 
At $z\approx1$, a larger fraction 
of passive galaxies follows the $z\approx0$ size -- mass relation. 
Superdense relics 
with $R_e \approx$ 1 kpc are extremely rare at $z\approx0$ with respect 
to $z>1$, and absent if $R_e<1$ kpc. Because of the similar sizes and 
mass densities, we suggest that the superdense passive galaxies at 
$1<z<2$ are the remnants of the powerful starbursts occurring in 
submillimeter--selected galaxies at $z>2$. The results are compared 
with theoretical models and the main implications discussed in the framework 
of massive galaxy formation and evolution. 
}
{}

\keywords{Cosmology: observations-- Galaxies: distances and
redshifts -- Galaxies: elliptical and lenticular, cD -- 
Galaxies: evolution -- Galaxies: formation -- Galaxies: high-redshift
}

\titlerunning{Superdense passive galaxies}

\maketitle

\section{Introduction}

Deep surveys provide the observational constraints needed
to understand galaxy formation and evolution. In particular, 
many studies have been focused on massive galaxies (i.e. stellar 
mass ${\cal M} > 10^{11}$ M$_{\odot}$) as cosmological probes of the 
history of galaxy mass assembly. However, despite the remarkable success 
in finding and studying massive galaxies over a wide range of cosmic time, 
the global picture is far from being clear. Thanks to their simple and 
homogeneous properties (morphology, colors, passively evolving 
stellar populations, scaling relations) and being the most massive 
galaxies in the present-day Universe, early-type galaxies (ETGs) are 
crucial to investigate the cosmic history of massive galaxies (e.g.
\cite{alvio} and references therein). 

At $z<1$, the most recent results seem 
now to agree in indicating that the majority of massive ETGs 
(${\cal M} > 10^{11}$ M$_{\odot}$) were already in place at z$\approx$
0.7-0.8, with a number density consistent with the one at z=0, 
whereas the evolution is more pronounced for the lower mass ETGs 
(e.g. \cite{fon04,yamada,bundy06,cdr06,borch,scarlata,brown,bundy07}). 
This mass--dependent evolutionary scenario, 
known as "downsizing" (\cite{cowie,gavazzi}), was proposed to 
explain the galaxy star formation histories, i.e. with massive galaxies 
forming their stars earlier and faster than the low mass ones. Recent 
results suggest that the downsizing concept should extend also to the 
stellar mass assembly evolution itself, i.e. with massive galaxies 
assemblying their mass earlier (e.g. \cite{cdr06,bundy06,franceschini,
bundy07,perez}), thus providing new and stringent constraints for the 
current models of galaxy formation (e.g. \cite{delucia}). 

The above results trace the evolution of the number density, 
luminosity and mass, but do not explain what is the mechanism with 
which ETGs assemble their mass and shape their morphology. 
Dissipationless ETG--ETG major merging (also called "dry" merging) 
has been advocated as an important mechanism to build up the masses 
of ETGs at $0<z<1$ (\cite{bell06,vd05}). However, this scenario 
seems difficult to be reconciled with the properties of the small-scale 
clustering of ETGs at $0.16<z<0.36$ (\cite{masjedi,masjedi2}), and (for massive 
ETGs) with the weak evolution of the high--mass end of the stellar 
mass function (e.g. \cite{cdr06,bundy06,bundy07,pozz07}). An alternative
process proposed recently to explain the assembly of ETGs is based
on multiple, frequent minor mergers at moderate redshifts (\cite{bournaud}). 

Several studies suggest that the critical redshift range where the
strongest evolution and assembly took place is at $1 \lesssim z \lesssim2$
(e.g. \cite{fon04,glaze,abraham07,arnouts}). However, the picture 
at these redshifts is even 
more controversial than at $z<1$, partly because of the observational 
difficulty to identify spectroscopically and study large samples of 
ETGs at $z>1$. The few ETGs 
identified spectroscopically so far up to $z\approx 2.5$ are very red 
($R-K>5-6$), dominated by passively evolving old stars with ages of 
1-4 Gyr, have stellar masses typically ${\cal M} \gtrsim 
10^{11} M_{\odot}$, and are strongly clustered with $r_0 \approx$ 8--10 Mpc
(\cite{cim04,glaze,mcc04,daddi05a,sar05,kong,kriek}). Daddi et al. (2005)
were the first to realize that a large fraction of these ETGs have 
smaller sizes 
($R_e \lesssim 1$ kpc) (see also \cite{cassata05}) and higher mass 
internal densities than present--day ETGs. This result was soon confirmed 
by other observations (\cite{trujillo06,zirm07,longhetti07,toft07,
trujillo07}). However, it is still unclear how to explain 
such size -- mass -- density properties in the context of ETG evolution.

The existence of a substantial population of old, massive, passively
evolving ETGs up to $z\approx 2$ was not predicted in galaxy 
formation models available in 2004-2005, and opened the question 
on how it was possible to assemble such systems when the Universe was 
so young. A promising mechanism which can provide a better agreement 
with the observations seems to be the "quenching" of the  star formation 
at high redshifts with AGN "feedback" (e.g. \cite{gra,menci06}).

The stellar ages and masses of the passive ETGs at $z\approx1-2$
require precursors characterized by strong ($>100$ M$_{\odot}$/yr) and 
short-lived (0.1-0.3 Gyr) starbursts occurring at $z > 2-3$. In addition,
such precursors should also have a large clustering correlation length 
$r_0$ comparable to that of passive galaxies at lower redshifts ($z<2$) 
(e.g. \cite{daddi00,firth02,kong,farrah06}), and compatible with
that expected
in the $\Lambda$CDM models for galaxies located in massive dark 
matter halos and strongly biased environments. Examples of 
precursor candidates have been found amongst starburst galaxies selected 
at $z>2$ with a variety of techniques (e.g. $BzK$, \cite{daddi04}; 
submm/mm, \cite{chapman}; ``Distant Red Galaxies'', \cite{franx}; 
optically-selected systems with high luminosity, \cite{sha}; IRAC 
Extremely Red Objects, IEROs, \cite{yan}; HyperEROs, \cite{totani}, and 
ULIRGs at $z\sim 1-3$ selected with {\it Spitzer Space Telescope} 
\cite{berta07}). 
Deep integral-field near-IR spectroscopy is being used to perform 
detailed studies of these precursor candidates in order to understand 
what are the main mechanisms capable to assemble massive galaxies 
with short timescales (e.g. \cite{swin,nfs06,wright,law}). To date,
the most detailed case is represented by a $BzK$--selected star-forming 
galaxy at $z=2.38$ which shows a massive rotating disk with high
velocity dispersion which may become unstable and lead to the rapid 
formation of a massive spheroid (\cite{genzel}).

In this paper, we exploit the combination of GMASS ultradeep spectroscopy,
HST imaging and optical -- to -- mid-infrared photometry to study the 
global properties of a new sample of 
passive galaxies at $1.3<z<2.0$ in order to place new constraints
on massive galaxy formation and evolution. This paper refers to other 
papers where more details can be found on the GMASS project 
(Kurk et al. 2007a in preparation), the large scale structure at
$z=1.61$ (Kurk et al. 2007b in preparation), the photometric Spectral Energy 
Distribution (SED) fitting analysis (Pozzetti et al. 2007 in 
preparation) and the morphological analysis (Cassata et al. 2007 in 
preparation). We adopt $H_0=70$ km s$^{-1}$ Mpc$^{-1}$, $\Omega_{\rm m}=
0.3$ and $\Omega_{\Lambda}=0.7$, give magnitudes in AB photometric system,
and assume a Chabrier (2003) Initial Mass Function (IMF).

\section{The GMASS sample}

GMASS ({\it ``Galaxy Mass Assembly ultra-deep Spectroscopic Survey''}
\footnote{http://www.arcetri.astro.it/$\sim$cimatti/
gmass/gmass.html}) is a project based on an ESO VLT Large Program
(Kurk et al. 2007a). 

The GMASS main scientific driver is to investigate the physical and 
evolutionary processes of galaxy mass assembly in the 
redshift range of $1.5<z<3$, i.e. in the epoch when the crucial
processes of massive galaxy formation took place. Photometric redshifts 
are not sufficient to fully address the above questions because they 
provide limited clues on the physical and evolutionary status of the 
observed galaxies. Spectroscopy is therefore essential to 
derive reliable spectroscopic redshifts, to perform detailed spectral and 
photometric SED fitting (with known spectroscopic redshift), and to 
characterize the nature and diversity of galaxies at $1.5<z<3$. 

The first step was to select a sample at
$4.5\mu$m by using the GOODS-South public image taken at that wavelength
with the \emph{Spitzer Space Telescope} equipped with IRAC (Dickinson
et al., in preparation). Firstly, we defined a region of $6.8\times6.8$ 
arcmin$^2$ (matching the field of view of FORS2) located within the 
GOODS-South field\footnote{http://www.stsci.edu/science/goods}, and 
covering $\sim$80\% of the \emph{Hubble Ultra Deep Field} (HUDF) 
(\cite{beckwith}). Secondly, the sample was selected by extracting 
{\sl all} the sources present in that field and having $m_{4.5}<23.0$ (AB) 
(corresponging to a flux density $>$2.3\,$\mu$Jy). This selection
provided 1094 objects belonging to what we call the {\sl GMASS sample},
which is clearly a pure flux-limited sample. Extended simulations showed 
that at $m_{4.5}<23.0$, the photometric completeness is $\sim$85\% (M. 
Dickinson and G. Rodighiero, private communications).

\begin{figure}
\centering
\includegraphics[width=\columnwidth]{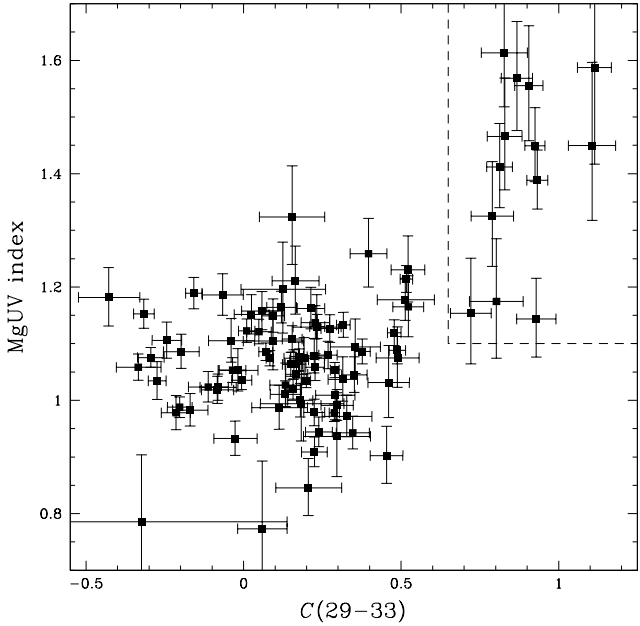}
\caption{The Mg$_{\rm UV}$ index vs. the UV color index defined in the 
text, for the GMASS galaxies with $z>1$. All the selected high-z passive 
galaxies fall within the box in the upper-left corner of the plot, having a 
distinctively red UV color index. Also, all but three galaxies show a
Mg$_{\rm UV}$ index greater than 1.2, typical of an evolved stellar 
population (see Daddi et al. 2005).
}
\end{figure}

The GMASS sample was then used to extract a sub-sample of target
galaxies to be observed spectroscopically. To reach the GMASS scientific 
aims, spectroscopy was very deep in order to derive secure 
spectroscopic redshifts for the faintest galaxies and to obtain 
high--quality spectra for the brighter galaxies in order to allow 
detailed spectral studies. The GMASS optical multi--slit 
spectroscopy was done with the ESO VLT + FORS2 (MXU mode) and 
focused on galaxies pre-selected with a cut in photometric redshift of 
$z_{phot}>1.4$ in order to concentrate the study on galaxies in the 
critical range of $1.5<z<3$. In order to make the spectroscopy feasible, 
two cuts in the optical magnitudes were adopted: $B<26.5$ or $I< 26.5$ for
spectroscopy done in the blue (grism 300V) or in the red (grism 300I)
respectively. The integration times were very long (up to 32 hours per 
spectroscopic mask), and the spectroscopy was optimized by obtaining 
spectra in the blue (4000--6000 \AA, grism 300V) or in the red 
(6000--10000 \AA, grism 300I) depending on the colors and photometric 
SEDs of the targets. For both grisms, the slit width was always 1 
arcsec and the spectral resolution $\lambda / \Delta \lambda \approx$600.
Despite the faintness of the targets, GMASS spectroscopy provided an overall 
spectroscopic redshift success rate of about 85\% for the targeted galaxies. 

Overall, $\approx$50\% of the GMASS flux-limited sample ($m_{4.5}<23.0$) 
has available spectroscopic redshifts if we combine our own spectroscopy 
with the literature redshifts (http://archive.eso.org/archive/adp/GOODS),
whereas the other half has accurate photometric redshifts (see Kurk et al. 
2007a).

In summary, the GMASS sample is a pure flux--limited sample selected
with $m_{4.5}<23.0$, and with spectroscopic (GMASS or literature) or 
photometric redshifts available for each object in the sample.
The power and the novelty of the GMASS sample is the selection at 
4.5$\mu$m, which is crucial for two main reasons: (1) the peak of the 
stellar SEDs ($\lambda_{rest}=1.6\mu$m) is redshifted in the 4.5$\mu$m 
band for $z>1.5$, (2) it is sensitive to the rest-frame near-IR emission, 
i.e. to stellar mass, up to z$\approx$3. For $m_{4.5}<23.0$ and a Chabrier 
IMF, the limiting stellar mass is log$({\cal M}$/M$_\odot) \approx$ 9.8, 10.1, 
and 10.5 for $z\approx1.4$, $z\approx2$, and $z\approx3$, 
respectively. This allows to properly investigate the galaxy mass 
assembly evolution within a wide range of masses. 

\section{Passive galaxies in the GMASS sample}

\subsection{Spectroscopic selection and properties}

At each redshift, the oldest envelope of the galaxy population
is a ``fossil'' tracer of past events of galaxy formation in a 
way complementary to the study of galaxies with ongoing star formation. 
ETGs can be used as "fossil" probes as they are usually dominated by 
old, passively evolving stellar populations.

The rest-frame UV spectra of old and passive stellar populations have
specific features different from those of younger or star-forming
galaxies. The UV continuum is very red, characterized by prominent 
breaks at 2640 \AA~ and 2900 \AA~ and rich of metal absorptions (e.g.
MgI$\lambda$2852, MgII$\lambda$2800, FeI and FeII lines) (e.g.
\cite{dunlop,spinrad,cim04,sar05,daddi05a}). The combination of the 
strongest breaks and absorptions at 2600--2900~\AA~ produces the 
prominent Mg$_{\rm UV}$ feature (\cite{daddi05a}).
The presence of this feature in a galaxy spectrum ensures that
very little if any star formation has taken place over the last $\sim
0.5$ Gyr, and therefore the galaxy is in a passively
evolving stage.  This feature is also very important to measure the
spectroscopic redshifts of this kind of galaxies using optical
spectrographs, when for $z> 1.4$ the more prominent features of ETGs
such as the CaII H\&K doublet and the Balmer/4000 \AA\ breaks fall
outside the optical spectral range.

\begin{figure*}
\centering
\includegraphics[width=\columnwidth]{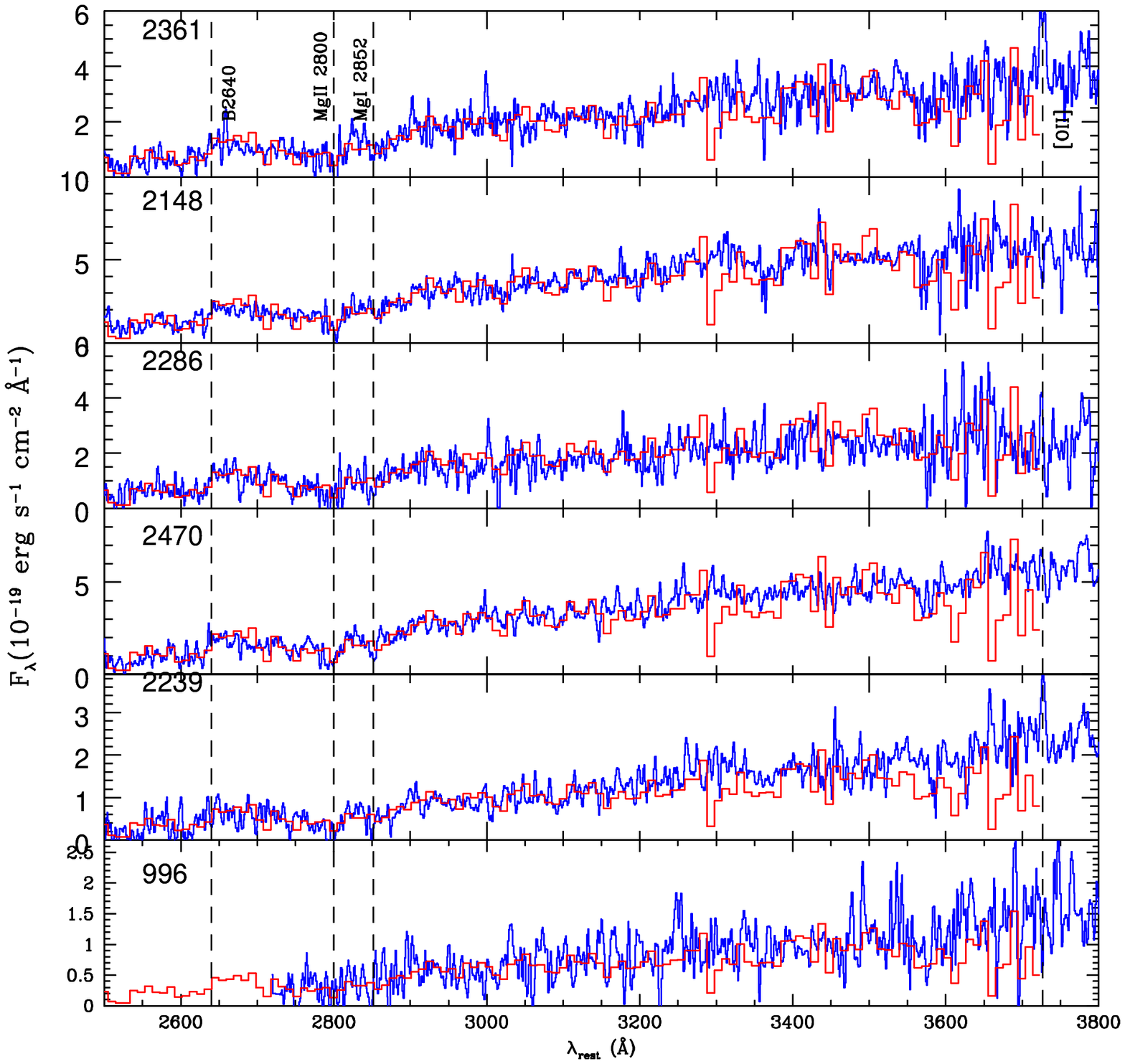}
\includegraphics[width=\columnwidth]{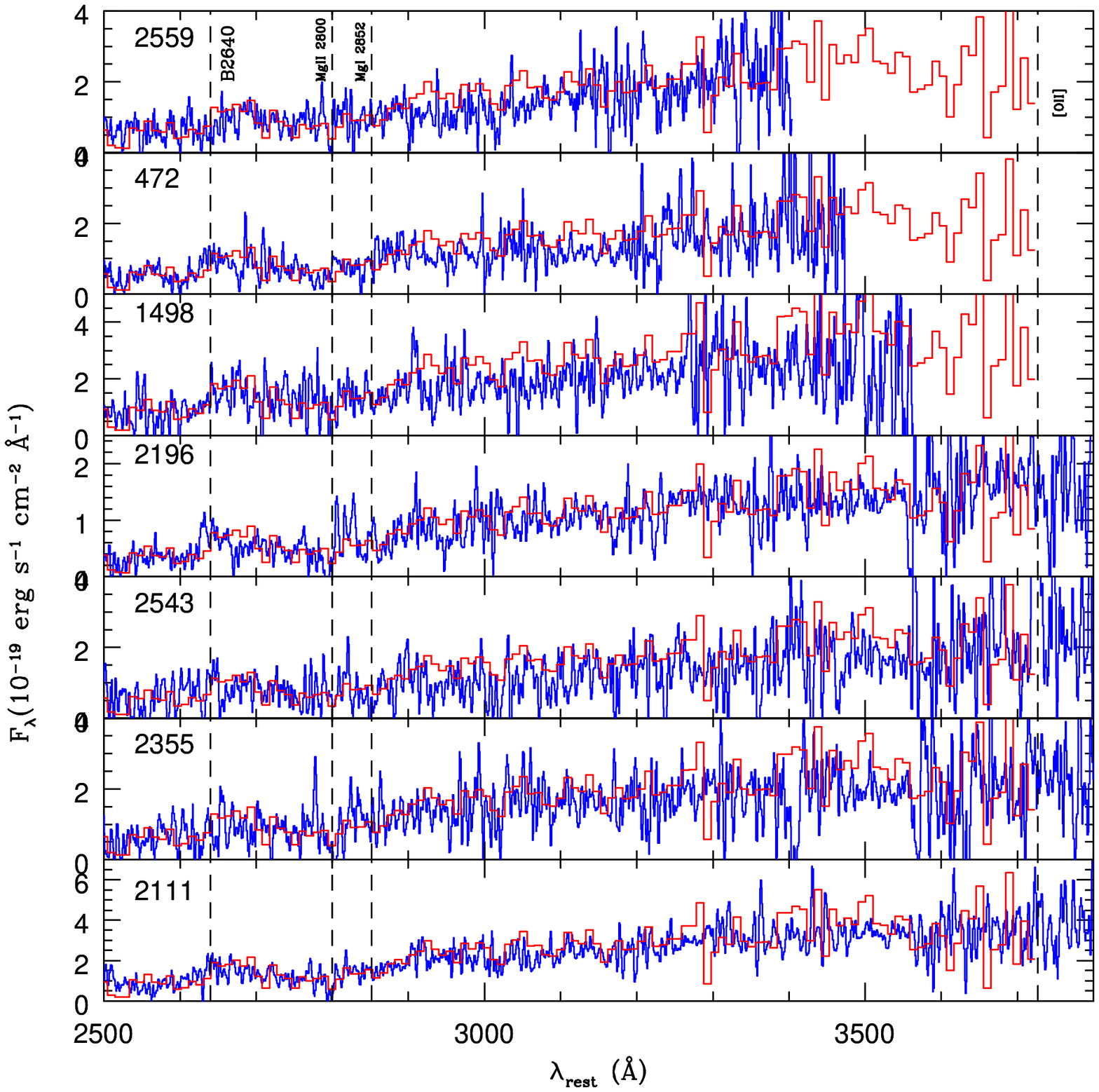}
\caption{The individual spectra of the selected galaxies ordered with 
increasing redshift from bottom to top in each panel. The red line is 
the spectrum of the old galaxy LBDS 53w091 ($z=1.55$; \cite{dunlop,spinrad}) 
used to search for spectra with a similar continuum shape.
}
\end{figure*}

Following this approach, the GMASS spectroscopic dataset was used to
search for passively evolving galaxies at $z>1.4$ with the Mg$_{\rm UV}$ 
feature present in their spectra (Mignoli et al., in preparation). 
This was achieved by deriving a color 
index of the UV continuum of each GMASS galaxy with available spectrum
and spectroscopic redshift $z>1$, and searching for the red continuum spectra
expected in the case of passive galaxies. This UV color was defined as :

$$ C(29-33) = -2.5log[F_{\nu}(2900)/F_{\nu}(3300)] $$

where $F_{\nu}(2900)$ and $F_{\nu}(3300)$ are the mean spectral 
continuum flux densities within the rest-frame wavelength bins at
2700-3100~\AA~ and 3100-3500~\AA~ respectively. The continuum flux densities
were measured in fixed rest-frame spectral ranges and adopting a {\sl
sigma-clipping} procedure which ensures that spikes due to bad sky 
subtraction and/or to cosmic rays residuals do not affect the measured 
values. The errors were then computed from the standard deviations.

Fig. 1 shows that a strong color segregation is present, with 13
galaxies at $1.39<z<1.99$ having $C(29-33)$ significantly redder than 
the rest of the GMASS spectroscopic sample. These red galaxies have also 
a Mg$_{\rm UV}$ index in the range expected for passive galaxies 
(\cite{daddi05a}). The visual inspection of these 13 spectra confirmed
their similarity with the spectra of LBDS 53w091 (see Fig. 1) 
(\cite{dunlop,spinrad}) and of old passive galaxies at $z>1$ taken from 
the K20 and GDDS surveys (\cite{mignoli05,cim04,mcc04}). All the other
GMASS spectra of galaxies at $z>1$ were visually inspected in order to
check for other passive/red galaxies with spectra that may have missed 
by the UV color criterion, but no other cases were found. 
The question of how many other red/passive ETGs are present in the {\sl
whole} GMASS flux-limited sample (i.e. including also galaxies with
only photometric redshifts) is beyond the scope of the present work and
is being addressed in another paper (Cassata et al. 2007, in preparation).

The spectra of the 13 selected galaxies are shown in Fig. 2 and some properties
are listed in Table 1. Seven of these galaxies belong to a large scale
structure at $z=1.61$ that is discussed in a companion paper (Kurk et al. 
2007b). 
Ten spectra were obtained with an integration time of 30 hours each,
whereas three (IDs 2111, 2239, 2559) were repeated in two independent
masks and each of them accumulated an integration time of 60 hours.
The two galaxies located within the HUDF (IDs 472 and 996) were also
consistently identified by Daddi et al. (2005a) (their IDs 3650 and 8238
respectively) using HST+ACS slitless grism spectroscopy (GRAPES project;
Pirzkal et al. 2004).

\begin{figure}
\centering
\includegraphics[width=\columnwidth]{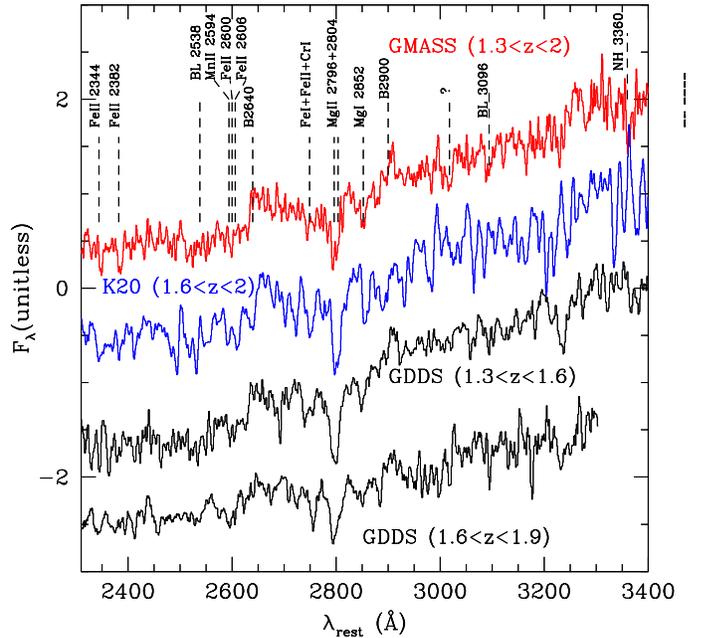}
\caption{Comparison of rest-frame UV spectra of passively evolving
galaxies from GMASS (this work), K20 (\cite{cim04}) and GDDS (\cite{mcc04})
surveys.}
\end{figure}

\begin{figure*}
\centering
\includegraphics[width=19cm,height=19cm]{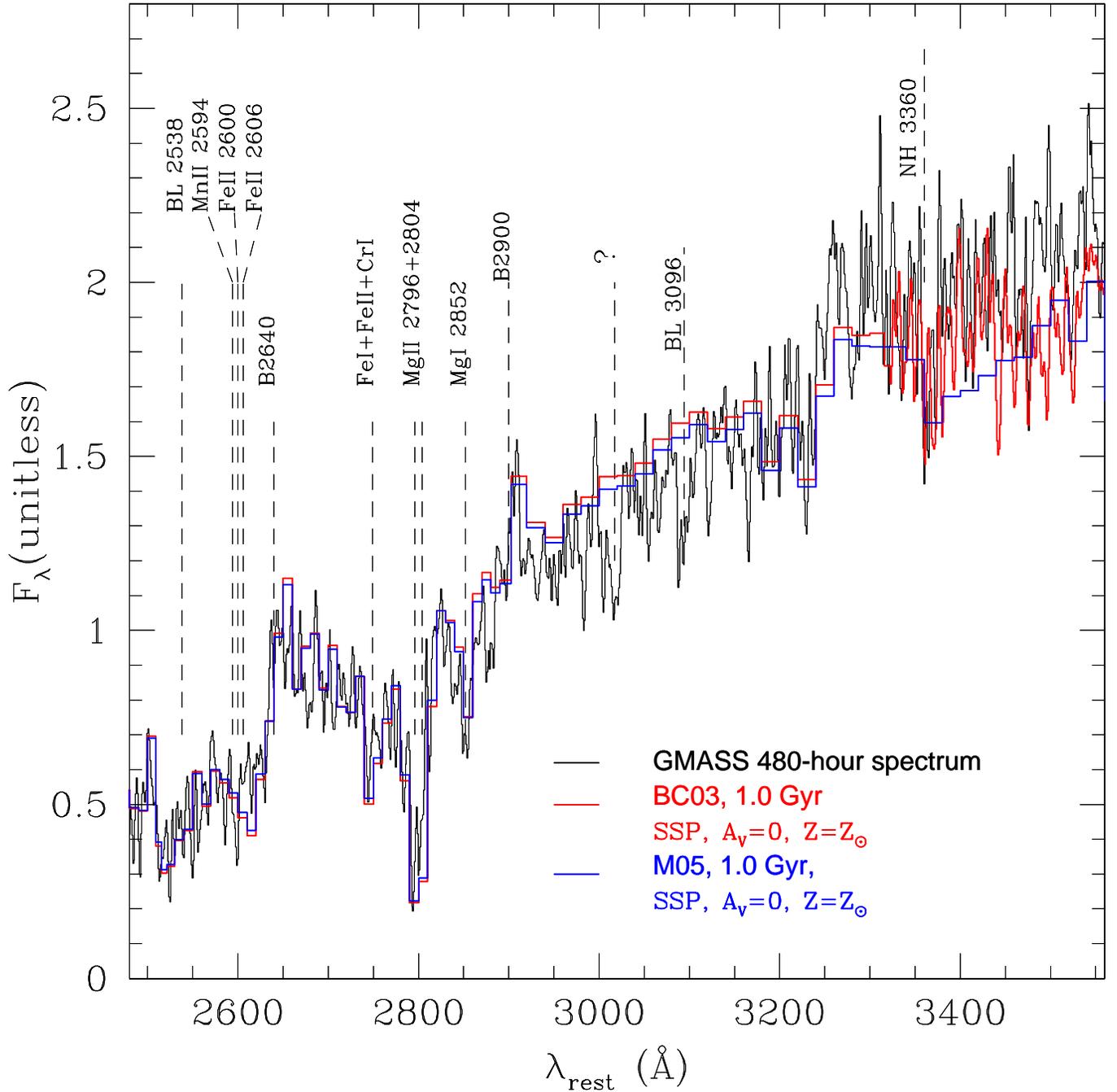}
\caption{The average spectrum of GMASS passive galaxies 
(black) and two synthetic spectra which provide the formal best 
fit in the range of 2480--3560~\AA~ for solar metallicity and age of 
1.0 Gyr for all the models (BC03 and M05). The CB07 best fitting
spectrum is not shown because it is identical to the BC03 one. The
question mark indicates an unidentified absorption feature at
$\sim$3018~\AA.
The stacked spectrum is publicly available at \emph{http://www.
arcetri.astro.it/$\sim$cimatti/gmass/gmass.html}.
}
\end{figure*}

Six spectra show no signs of star formation due to the lack of [O
II]$\lambda$3727 emission, four have weak [O II] emission, and
for the three galaxies at $z>1.8$ (IDs 472, 1498, 2559) there is no
information on [OII]$\lambda$3727 because the line falls outside the
observed spectral range. Assuming that the [O II] emission line fluxes 
(Table 1) are not due to an AGN, and using the conversion from 
[OII]$\lambda$3727 luminosity and
star formation rate (SFR) of Kennicutt (1998), we derive SFR$\approx$
1--4 M$_{\odot}$yr$^{-1}$ for the four galaxies with
[OII]$\lambda$3727 emission (see Table 1).  If dust extinction is
present, these SFRs are lower limits.  However, the similarity of the
spectra of these four galaxies with the spectra of galaxies
without line emission suggests that the bulk of the stars is made by
old stellar population and that the [OII]$\lambda$3727 emission is
probably due to a minor episode of star formation (see also Section
5.2 on the specific star formation rate), or to weak nuclear activity.

\begin{table*}
\caption{The sample of passive galaxies}   % title of Table
\label{table:1}      % is used to refer this table in the text
\centering                          % used for centering table
\begin{tabular}{c c c c c c c c c c}        % centered columns (4 columns)
\hline\hline                 % inserts double horizontal lines
ID &  Coordinates (J2000) & m$_{4.5}$ & $I$ & $K_s$ & $B-$z  & z$-K_s$ & spec-$z$ & F$_{obs}$([O II]$\lambda$3727) & EW$_{obs}$([OII]$\lambda$3727) \\
& & & & & & & & $\times 10^{-18}$ erg s$^{-1}$ cm$^{-2}$ \AA$^{-1}$ & \AA \\

\hline                        % inserts single horizontal line
472&53.1588440 -27.7971249&21.04&25.29&21.72&3.25&2.57&1.921&---- & ---- \\
996&53.1538162 -27.7745972&21.36&25.27&22.02&3.57&2.27&1.390&$<1$ & $<7$ \\
1498&53.1746445 -27.7533722&20.72&25.35&21.57&2.54&2.80&1.848&---- & ---- \\
2111&53.1164055 -27.7126999&20.64&24.58&21.31&3.99&2.52&1.610&$<2$ & $<6$ \\
2148&53.1512375 -27.7137375&19.80&24.17&20.50&4.57&2.71&1.609&$<2$ & $<5$\\
2196&53.1527710 -27.7162437&20.34&24.53&20.92&4.28&2.51&1.614&$<2$ & $<10$ \\
2239&53.1304817 -27.7211590&20.61&24.53&21.24&4.08&2.39&1.415&4.0$\pm$0.5 & 22$\pm$4 \\
2286&53.1249580 -27.7229443&20.82&25.08&21.55&3.69&2.53&1.604&3.5$\pm$0.5 & 29$\pm$5 \\
2355&53.0596466 -27.7258034&20.93&24.83&21.78&2.90&2.30&1.610&$<2$ & $<9$\\
2361&53.1085548 -27.7101574&20.40&24.90&21.10&4.08&2.71&1.609&8.0$\pm$1.0 & 26$\pm$4 \\
2470&53.1421547 -27.7112675&20.16&23.70&20.68&3.85&2.14&1.416&$<2$ & $<4$\\
2543&53.1496468 -27.7113838&20.26&25.24&21.25&3.40&3.20&1.612&15.0$\pm$5.0 & 100$\pm$40 \\
2559&53.1764069 -27.7011585&20.71&24.84&21.60&3.37&2.81&1.981&---- & ---- \\
\hline                                   %inserts single line
\end{tabular}
\end{table*}

While the presence of Type 1 AGN is excluded by the lack of broad
emission lines (e.g. MgII$\lambda$2800), we cannot rule out the
presence of hidden AGN activity. For instance, Yan et al. (2006)
found that a large
fraction of SDSS galaxies at $z<0.1$ with red rest-frame colors and
emission lines display line ratios typical of various types of AGNs
(LINERs, transition objects, and, more rarely, Seyferts) rather than
of star-forming galaxies. In the case of our spectra, emission line
ratio studies are not possible because the relevant lines fall
outside the observed range. This implies that the above SFRs could be
taken as upper limits. However, the lack of high ionization lines
(e.g. [NeV]$\lambda$3426) makes the AGN hypothesis unlikely. The
presence of hidden luminous AGNs is also excluded by the stacking
analysis of the X-ray data (see Section 4). Therefore, we conclude
that the [OII] emission in these 4 galaxies is likely due to a
very low level of ongoing star formation. 

In the present paper, we focus the analysis on these 13 galaxies for
which ultradeep GMASS spectra are available. We only mention here 
that other passive galaxy \emph{candidates} are present in the GMASS sample
in the same redshift range. However, no spectroscopic redshifts are
available for these candidates and their properties are based only 
on the similarity of their SED fitting parameters with those of the 
13 spectroscopic passive galaxies discussed here (see Section 5.2). 
The detailed study of these candidates will be the subject of a 
forthcoming paper (Cassata et al. 2007 in preparation).

\subsection{Analysis of the 480--hour stacked spectrum}

The availability of deep individual spectra provides the opportunity
to add them together and to obtain a \emph{stacked} spectrum
with unprecedented depth and 
suitable for a detailed analysis. We emphasize that the GMASS stacked
spectrum of the 13 ETGs has an equivalent integration time of
(10$\times$30h) + (3$\times$60h) = \emph{480 hours} (1.73 Ms). The 
need for such extremely long integration times highlights the limits 
of the current
generation of 8-10m telescopes in obtaining spectra of red passive
galaxies with faint optical magnitudes down to $I \approx$25--26.
Moreover, near-infrared continuum and absorption line
spectroscopy of this kind of galaxies with $K_s \gtrsim 21$
is beyond the current limits of 8-10m telescopes. Thus, these
galaxies are ideal targets for the next generation of large
telescopes on the ground (e.g. E-ELT, TMT) and in space (JWST).

The stacked spectrum shown in Fig. 3--4 was obtained by averaging all
the 13 individual spectra (average redshift $\langle z \rangle=1.6$). 
In particular,
each spectrum was blue-shifted to the rest-frame according to its
redshift (with a 1.0~\AA~ \ rest-frame bin) and normalized in the
2600-3100~\AA~ \ wavelength range, which is always covered in the
observed spectral window. In order to avoid any bias towards the
brightest galaxies with the highest S/N ratio, the same weight was
assigned to each individual spectrum.
The main features present in the stacked spectrum are listed in Table 2.
The stacked spectrum is publicly available at \emph{http://www.
arcetri.astro.it/$\sim$cimatti/gmass/gmass.html}.

\begin{figure}
\centering
\includegraphics[width=\columnwidth]{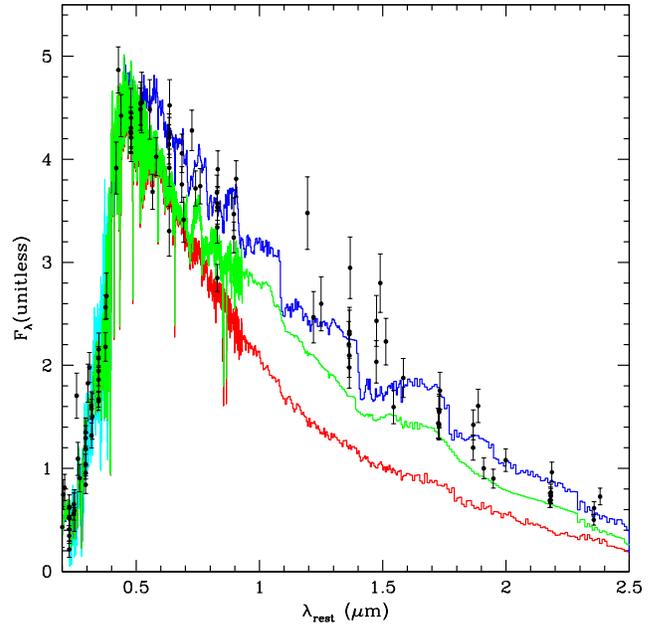}
\caption{Same as in Fig. 4, but including also the CB07
best fitting spectrum (green) and showing how the templates differ with
each other at longer wavelengths. The observed stacked spectrum
is shown in cyan. The black dots are the rest--frame photometry 
of the galaxies normalized at $\lambda_{\rm rest}=0.5\mu$m (see 
text for more details).}
\end{figure}

Fig 3 shows a comparison between the average spectrum of all
13 GMASS ETGs and those of K20 (\cite{cim04}) and
GDDS (\cite{mcc04}) samples. With the exception of the composite GDDS
spectrum
at 1.6$<z<$1.9 (slightly bluer and with less pronounced
Mg$_{\rm UV}$ feature), the spectra are very similar to each other
both in the strength of the Mg-UV feature and continuum slope.

\begin{table}
\caption{Spectral features in the stacked spectrum}   % title of Table
\label{table:2}      % is used to refer this table in the text
\centering                          % used for centering table
\begin{tabular}{l l c l}        % centered columns (4 columns)
\hline\hline                 % inserts double horizontal lines
Feature& Value & Type & Reference \\
\hline                        % inserts single horizontal line
FeII 2344 & 4.5$\pm$0.3 \AA & Equivalent Width & Ponder et al. 1998 \\
FeII 2382 & 6.0$\pm$0.4 \AA & Equivalent Width & Ponder et al. 1998 \\
BL 2538   & 5.0$\pm$1.0 \AA & Equivalent Width & Ponder et al. 1998 \\
B2640     & 1.59$\pm$0.06 & Continuum Break & Spinrad et al. 1997\\
B2900     & 1.27$\pm$0.03 & Continuum Break & Spinrad et al. 1997\\
MgII 2800 & 8.0$\pm$0.6 \AA & Equivalent Width & Ponder et al. 1998 \\
MgI 2852  & 8.0$\pm$0.6 \AA & Equivalent Width & Ponder et al. 1998\\
BL 3096   & 2.5$\pm$0.5 \AA & Equivalent Width & Ponder et al. 1998\\
NH 3360   & 2.5$\pm$0.4 \AA & Equivalent Width & Ponder et al. 1998\\
Mg$_{\rm UV}$ & 1.40$\pm$0.1 & Continuum Break & Daddi et al. 2005\\
\hline                                   %inserts single line
\end{tabular}
\end{table}

The observed average spectrum was compared to various libraries of
synthetic spectra of simple stellar populations (SSPs) in order to
attempt an estimate of the age of the stellar population dominating
the rest-frame UV spectrum covered by our observations. In practice,
limiting the analysis to this part of the spectrum provides an
estimate of the time elapsed since the last major episode of star
formation. As such, this estimate must be regarded as a {\it lower}
limit to the age of the bulk of the stars in these galaxies. As a 
second step we included also the rest-frame near-IR, so as to have 
an estimate of the age of the bulk stellar populations in
these galaxies (see Section 5.2). 

For the fitting analysis, we adopted three libraries of synthetic
spectra: Maraston (2005; M05) (Kroupa 2001 IMF), Bruzual \& Charlot;
BC03) (Chabrier IMF), and its update (Charlot \& Bruzual 2007; CB07) 
(Chabrier IMF). The
main difference among the three sets of models is in the treatment of
the thermally pulsing asymptotic giant branch (TP-AGB) phase of
stellar evolution, which for intermediate age SSPs can contribute up
to $\sim 50$\% to the total bolometric light, radiated mostly in the
near-IR (Renzini 1981; Maraston 1998, 2005; Maraston et al. 2006). In
M05 models the TP-AGB contribution was calibrated using Magellanic
Cloud globular clusters of various ages, whereas BC03 models had a
negligibly small contribution of the TP-AGB at all ages (M05). With
CB07 an attempt was made to obviate to this deficiency (\cite{bruzual07}). 
We concentrate here on the analysis of the rest-frame UV spectrum alone, 
for which the
TP-AGB contribution is irrelevant, postponing to Section 5.2 the 
inclusion of the whole photometric information on these galaxies, 
including the rest-frame near-IR.

The BC03 and CB07 models have a spectral resolution of $\lambda / \Delta
\lambda \sim$300 (i.e. $\sim$10~\AA) for $\lambda<$3300~\AA~ and
$\Delta \lambda \sim$3~\AA~ for $\lambda>$3300~\AA, whereas the M05 
spectra have 
$\lambda / \Delta\lambda \sim$300 at all wavelengths. The observed stacked 
spectrum (resolution $\lambda / \Delta \lambda \sim$600, i.e.  $\Delta 
\lambda \sim$2~\AA~ at $\lambda_{rest}=$3000~\AA, $\langle z \rangle$=
1.6) was rebinned to match the resolution of the model spectra.

The best fit age for each set of synthetic templates was derived
through a $\chi^2$ minimization over the rest-frame wavelength range
2480--3560~\AA~ that is covered completely by 11 out of the 13 spectra.
The $rms$ as a function of wavelength used in the $\chi^2$
procedure was estimated from the average spectrum computing a running
mean $rms$ within 20~\AA~ and 10~\AA~ for $\lambda<$3300~\AA~ and
$\lambda>$3300~\AA~ respectively. All ages available in the model
spectra libraries were used, and we initially adopted $A_V=0$ during the 
fitting.

In the case of solar metallicity, the best fitting spectra have an age
of 1.0 Gyr for all models (BC03, M05 and CB07) (Fig. 4), indicating the 
three sets of models are virtually equivalent in this wavelength range
(BC03 and CB07 are actually identical, the only difference being at
longer wavelengths). 

In order to assess the relevance of the age--metallicity degeneracy, 
we explored how the estimated age changes as a function of metallicity 
$Z$ with BC03/CB07 models. This was done by assuming no dust extinction 
($A_V$=0) and evaluating how the best-fitting age changes as a function 
of $Z$ using six values of metallicity: 
$Z$=0.0004, 0.001, 0.004, 0.008, 0.019, 0.03. We found that the best-fit 
ages vary from 0.7 Gyr to 2.8 Gyr for $Z=$1.5--0.2 $Z_{\odot}$. 
Lower metallicities ($Z<0.2 Z_{\odot}$) 
are excluded because they never reproduce the observed spectrum. 
Although the results of the SED fitting (see Section 5.2) exclude the
presence of substantial dust extinction (not surprisingly, as most of
the sample galaxies show no or very weak star formation),
we also studied the effects of including the dust attenuation in the
fitting process. This was done by leaving the stellar age and the dust 
extinction ($A_V$) as free parameters for each fitting done with a fixed 
metallicity $Z$ (with $0.0004 \leq Z \leq 0.03$, with the same six 
metallicity values as above). The dust extinction 
was allowed to vary in the range of $0\leq A_V \leq 0.6$ in order to be
consistent with the results of the SED fitting of Section 5.2. Compared
to the case with no dust extinction ($A_V$=0), the best-fitting ages 
were found to slightly decrease to 0.6--2.3 Gyr for $Z=$1.5--0.2 
$Z_{\odot}$. This excercise clearly showed that 
the dominant degeneracy is the one of age and metallicity, which
become substantial when using the UV spectrum alone. 

We note that in all cases, the oldest acceptable ages 
never exceed the age of the Universe (3.27 Gyr) at the redshift 
($z=1.981$) of the most distant galaxy in the stacked spectrum.
Fig. 4 shows the best-fitting model spectra (for
$Z=Z_{\odot}$) and the observed stacked spectrum. 
However, Fig. 5 shows that the same model spectra which provide equally good
fits in the rest--frame UV differ significantly at longer wavelengths
(see also \cite{maraston06}). Fig. 5 shows also the "stacked" photometric 
SED obtained by de--redshifting the multi--band photometry ($BVIzJHK_s$
+3.6$\mu$m, 4.5$\mu$m, 5.8$\mu$m, 8$\mu$m) of each galaxy and normalizing it
to the average flux of the three SSP spectra at $\lambda_{\rm rest}=0.5\mu$m,
a wavelength at which all the three model spectra are nearly identical. 
The photometry of ID 2543 was excluded because significantly 
redder from the SEDs of the other 12 galaxies at $\lambda_{\rm 
rest}>1\mu$m. Fig. 5 shows that for the three different SSP 
spectra with a fixed age of 1 Gyr, the "stacked" photometric SED is in 
better agreement with the M05 model (although the scatter in the photometry
is significant). The same figure also clearly suggests that more stringent
constraints on the stellar population content of the passive galaxies
come from the photometric SED fitting analysis extended to the {\it 
Spitzer Space Telescope} IRAC bands. This is discussed in details
in Section 5.2.

\section{X-ray and 24$\mu$m emission}

In order to search for additional indicators of "activity" due to hidden 
star formation and/or AGN, we explored whether the 13 passive galaxies 
were detected in the X-ray or 24 $\mu$m data using the Chandra X-ray 
Observatory and the {\it Spitzer Space Telescope} + MIPS data 
publicly available for the GOODS-South/GMASS region.  

X-ray data for a total exposure time of $\sim$1 Msec obtained with the
\textit{Chandra} observatory in the GOODS/CDFS field have been
published and made publicly available by Giacconi et al. (2002, see
also Alexander et al. 2003). The limiting flux of the X-ray data
is $\sim 1\times10^{-16}$ erg cm$^{-2}$ s$^{-1}$ in the full band at
0.5--7 keV. We searched for 
individual emission from the 13 objects in the present sample by 
cross-correlating the IRAC positions of our targets with the positions 
of X-ray sources as catalogued by Giacconi et al. (2002) and Alexander 
et al. (2003). We found that none of the 13 sources was individually 
detected in the X-rays. In order to constrain the average X-ray 
properties of our sample galaxies, we used the ``stacking technique''
following Brusa et al. (2002). For the photometry, a circular 
aperture with a radius of $2''$ centered at the positions of our sources 
was adopted. The counts were stacked in the standard soft, hard and 
full bands (0.5-2 keV, 2-7 keV, and 0.5-7 keV) for a total effective 
exposure time of $\approx$ 11 Ms. Extensive Monte Carlo simulations 
(up to 10,000 trials) have been carried out by shuffling 13 random 
positions and using the same photometry aperture (2 arcsec). 
The random positions were chosen to lie in ``local background regions'' 
to reproduce the actual background as close as possible. In none of the
bands (full, hard, soft), a signal has been significantly detected.
The background counts at 0.5-2 keV imply a count rate of 
$< 5.2\times 10^{-6}$ counts s$^{-1}$ and a flux $< 3 \times10^{-17}$ erg 
cm$^{-2}$ s$^{-1}$ assuming an unobscured $\Gamma=2.0$ power-law 
spectrum. This implies upper limits on the rest-frame luminosity of 
$< 5.2 \times 10^{41}$~erg~s$^{-1}$ at 0.5-2 keV. These limits 
imply that a luminous AGN source ($L_X>10^{42}$ erg/s) is absent 
in the passive galaxies of our sample or is very heavily obscured.

Around $z\approx$1, about 20\% of the morphologically--selected
ETGs show some level of ``hidden'' activity revealed
through the mid-IR emission at 24$\mu$m (\cite{rodig07}).
Mid-infrared emission was searched using the public data available
for GOODS-South obtained with Spitzer + MIPS at 24 $\mu$m with 
an integration time of about 10 hours per sky pixel.
The photometry was based on a PSF fitting algorithm, where the
\emph{SExtractor} (\cite{bertin}) positions of the IRAC sources were 
used as input to the
MIPS source extraction process (Chary et al. in preparation;
see also \cite{daddi07a,daddi07b} for 
more details). None of the 13 galaxies is a strong MIPS source. 
The fluxes of 12 galaxies are not significant ($<2-3 \sigma$) and 
$F_{24}<$20 $\mu$Jy. The only exception is object ID 2543 which has a 
possible detection at 4.7$\sigma$ significance level with a flux 
$F(24)=24.7 \pm 5.3$ $\mu$Jy. We recall that ID 2543 is also the 
galaxy with the strongest [OII]$\lambda$3727 emission.
If the 24$\mu$m emission is real and due to star formation activity, 
and extrapolating from the 24$\mu$m measurements using the spectral
templates of Chary \& Elbaz (2001) (see for instance \cite{daddi07a,
daddi07b}, the implied infrared (bolometric) luminosity is $\approx 
1.4 \times 10^{11}$ L$_{\odot}$. If we convert
this luminosity into a SFR using the recipe of Kennicutt (1998), we
obtain $\approx$ 24 M$_{\odot}$ yr$^{-1}$, higher than the 
estimate based on [OII]$\lambda$3727 luminosity (4 M$_{\odot}$
yr$^{-1}$) or SED fitting (12 M$_{\odot}$ yr$^{-1}$ with BC03 models). 

\section{Photometric properties and SED fitting}

\subsection{Colors}

The location of the selected galaxies in the $I-K_s$ vs. $K_s-m_{4.5}$
two-color plot is displayed in Fig. 6, showning that these passive
galaxies are among the reddest in both colors if compared to the GMASS
galaxies with spectroscopic redshifts.
In particular, their colors are all broadly consistent with
the ones expected for passively evolving galaxies in the $BzK$
diagram (Fig. 7) (\cite{daddi04}), especially when allowance is made
for photometric errors. Note instead that these galaxies do not
qualify as ``Distant Red Galaxies'' (DRGs, \cite{franx}), because
they are all at $z<2$, whereas the DRG selection criterion
($J-K_s>2.3$ or $J-K_s>1.37$ in Vega or AB system) is designed
to select passive galaxies at $z>2$.  Indeed, the $J-K_s$
colors of our passive galaxies are all consistently $J-K_s \leq 1.37$
(actually, $0.7<J-K_s \leq 1.37$).

\begin{figure}
\centering
\includegraphics[width=\columnwidth]{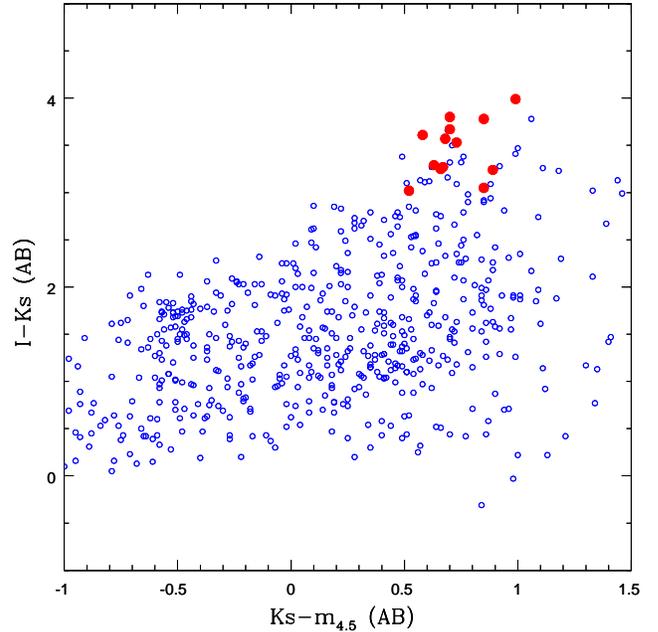}
\caption{The location of the selected passive galaxies (red
filled
circles) in the $I-K_s$ vs $K_s-m_{4.5}$ diagram. Open blue
symbols indicate GMASS galaxies with spectroscopic
redshifts.}
\end{figure}

\begin{figure}
\centering
\includegraphics[width=\columnwidth]{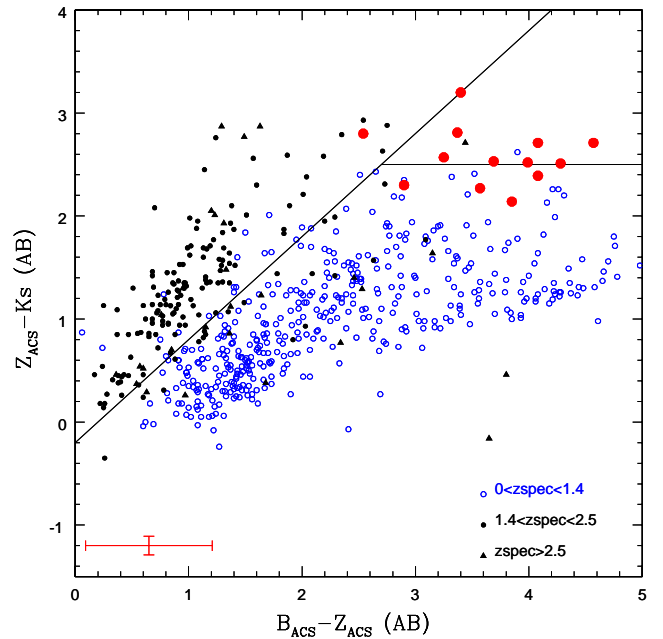}
\caption{The location of the selected passive galaxies (red
filled circles) in the $BzK$ color-color diagram (\cite{daddi04}).
Blue open circles, black filled circles and triangles indicate 
GMASS galaxies with known spectroscopic redshifts 
$z<1.4$, $1.4\leq z \leq 2.5$ and $z>2.5$ respectively.
The red error bars in the bottom left corner indicate
the average photometric uncertainty on the $z-K_s$ and $B-z$
colors for the selected passive galaxies. 
The errors on $B-z$ are very large due to the very faint $B$-band magnitudes.
}
\end{figure}

\subsection{SED fitting}

As discussed in Section 3.2 and in previous studies
(e.g. \cite{cotter, maraston06}), the rest-frame UV spectra are only
sensitive to the most recent star formation events, whereas including
in the analysis all the available photometric bands extending to the
rest-frame near-IR should allow us to gather information also on the
previous star formation history (SFH).

The photometric SED was derived for each of the 13 ETGs using the
public images available in the GOODS-South
field including 11 bands: optical (HST+ACS, $BVIz$; \cite{giava04}),
near-infrared (ESO VLT+ISAAC, $JHK_s$) and mid-infrared (Spitzer+IRAC,
3.6$\mu$m, 4.5$\mu$m, 5.8$\mu$m, 8$\mu$m, Dickinson et al. in
preparation).

We have used the synthetic spectra of M05 (Kroupa IMF) and BC03/CB07
(Chabrier IMF), adopting exponentially declining star formation
histories $SFR = ({\cal M}/ \tau) exp(-t/ \tau)$ with $\tau=$ 0.1,
0.3, 1, 2, 3, 5, 10, 15, and 30 Gyr, plus the case of constant star
formation rate. For each value of $\tau$, a set of synthetic SEDs is
pre-calculated for ages (since the beginning of the model at $t=0$) of
0.1, ..., 12 Gyr, in steps of 0.1, Gyr, and only $\tau$-models with
ages less than the age of the Universe at the redshift of each galaxy
are retained in the best-fit procedure. Extinction has been treated as
a free parameter in the optimization, having adopted the extinction
curve of Calzetti et al. (2000).
We have adopted solar metallicity for all the models. 
The fitting procedure selects the template spectrum that minimizes the
$\chi^2$, and therefore gives a value for each of four parameters: the
``age'' of the best-fitting model, the $e$-folding time of the SFR
$\tau$, the extinction $A_{\rm V}$, and the stellar mass.
This latter quantity comes from the absolute normalization of the
SED, once the best-fitting model has been chosen as the one providing
the closest SED shape to the observed SED.

We caution that an exponentially declining SFR is certainly a
convenient mathematical assumption, but the actual star formation
history may be significantly different. At least for models with small
$\tau$ values, the ``age'' given by the
best-fitting procedure must be interpreted as {\it the age
of the bulk stellar population of a galaxy}. Of course, real galaxies
are likely to contain also stars significantly older than the ``age''
derived in this way. On the other hand, the best fit ``age'' derived
from just the rest-frame UV spectrum (as in Section 3.2) corresponds to
the age of the last significant episode of star formation.

During the fitting, only the observed bands corresponding up to the
rest-frame $K_s$-band ($\lambda_{rest}< 2.5\mu$m) were used in order
to avoid any dust emission contamination. The redshifts are taken
from the GMASS spectroscopy and are {\it not} free parameters,
hence largely reducing the SED fitting degeneracies
often occurring when also the (photometric) redshift is a free
parameter.  The absence of luminous AGN sources as inferred from the
X-ray and 24 $\mu$m data (see previous Section) implies that the
observed flux measured from the sample galaxies is dominated by
stellar radiation. We have also explored the effect of assuming other
than solar metallicities, from sub-solar to super-solar. This was
attempted only using the CB03/BC07 models, but the best fit was always
achieved for solar metallicity. Table 3 shows the main results of SED
fitting in the case of solar metallicity.

\subsubsection{The Stellar Masses and Their Uncertainties}

The procedure adopted here for measuring the stellar masses is
basically the same as that generally followed in the literature
(e.g. Fontana et al.  2004, 2006; Longhetti et al. 2005; 
Bundy et al. 2005, 2006; Pozzetti et
al. 2007; Trujillo et al.  2007). While the details of our procedure are
extensively discussed in Pozzetti et al. (2007), we recall
here a few points. The internal accuracy of the measured stellar
masses is $\sim 0.2$ dex, as derived from the width of the probability
distributions in the fitting procedure. In addition to this typical internal
error, systematic errors should also be mentioned. A first source of
systematics is the choice of the IMF, which is not a free parameter in
our procedure, but must be adopted {\it a priori}. Stellar masses
derived assuming the Kroupa IMF are systematically higher by 0.04 dex
than those derived assuming the Chabrier IMF, and this difference was
subtracted from the masses derived with Maraston (2005) models (using Kroupa
IMF), in order to allow a homogeneous comparison with the masses
derived using the Bruzual \& Charlot (2003) and the Charlot \& Bruzual
(2007) models. The use of a straight, single-slope Salpeter IMF would
result in masses about 70\% higher. Another source of uncertainty is the
assumption made on the star formation history. For instance, detailed 
simulations (e.g. \cite{fon04,pozz07}) showed that in the case of an 
underlying exponentially declining SFH with $SFR \propto exp(-t/ \tau$)
plus random bursts of star formation superimposed on it, the estimated
stellar masses become higher by $\approx$30-40\%. Finally, as mentioned 
above the stellar population models differ in their treatment of the TP-AGB
phase of evolution, which is particularly important for stellar
populations in the age range $0.5-2$ Gyr, and most of the galaxies in
the present study fall in this range of ages. As already pointed out
by Maraston et al. (2006) and Wuyts et al. (2007), BC03 models give
stellar masses that are systematically larger by $\approx$40--60\%
than those estimated with M05 models (see Fig. 8). 
This is most likely due to the
best fitting algorithm boosting the mass in order to compensate for the
lacking TP-AGB contribution in the BC03 models. The
stellar masses from CB07 and M05 models are instead in fairly good
agreement. It is important to remind that similar uncertainties 
on photometric stellar masses are also present in the SED fitting 
of low-$z$ SDSS galaxies (see \cite{kauff03}). However, we also recall 
that the stellar masses are generally in good agreement with the 
dynamical masses, especially for early-type galaxies 
(see e.g. \cite{drory04,ssa,rettura}).

The stellar masses (and luminosities) of our sample galaxies 
span a rather wide range of values, from very massive and
luminous (like the high-$z$ ETGs discussed in \cite{cim04,mcc04}), to
a regime of lower luminosity and stellar masses that was possible to
reach thanks to the deep GMASS spectroscopy (-20.4$<M_B <$-22.5,
$10.0<$log(${\cal M}$/M$_{\odot})<$11).)

\subsubsection{The Age of the Dominant Stellar Population}

In the case of M05 models the ages of the dominant stellar populations
are remarkably homogeneous around $\sim 1$ Gyr, with a small scatter
of $\pm 0.2$ Gyr (Tab. 3). The best fit is always achieved with the smallest
$e$-folding time, i.e., $\tau=0.1$ Gyr. Being $\langle z \rangle=1.6$ for these
galaxies, this age estimate implies that the bulk of the stellar
populations should have formed at $z=2-2.5$. The same procedure
applied to BC03 and CB07 models gives a somewhat older average age,
$\sim 1.6$ Gyr, implying a peak formation redshift $\sim
2.5-3$. However, in this case there appears to be a much larger age
dispersion ($\pm 0.8$ Gyr), with some ages being as high as $\sim 3$
Gyr, corresponding to a peak formation redshift $\sim 5$. In general,
for these {\it older} galaxies the $e$-folding time indicated by CB07
models is longer, and the stellar mass somewhat higher than indicated
by M05 models.  This discrepancy suggests that the two sets of models
are still appreciably different, most likely in the way the TP-AGB
phase is implemented, as it is clearly shown in Fig. 5. 

We note that for all galaxies when using M05 models, and for most using
the CB07 ones, the SED ages derived in this section are very much
consistent with that derived solely from the rest-frame, stacked UV 
spectrum in Section 3, i.e. $\sim 1$ Gyr. We interpret this agreement 
as an additional indication that most of the star formation activity was 
confined within a short time interval, short compared to the typical age 
($\sim 1$ Gyr) of the stellar populations. In this respect, we also note 
that in spite of $\tau$ models having been widely used in the literature, 
nature may have proceeded differently: rather then starting abruptly at 
its maximum intensity and then declining exponentially, star formation is 
likely to have started at low level at very high redshift, then increasing 
exponentially as mass is accumulated with $\dot {\cal M} \propto {\cal M}$ 
(Daddi et 
al. 2007a), and finally suddenly truncated by feedback effects. If so, an 
{\it inverted $\tau$ model} would be more appropriate.

We can conclude that, by and large,
our analysis indicates that the bulk of stars in these passively
evolving galaxies must have formed at $z \approx 2-3$, a result in excellent
agreement with the evidence for ETGs from $z=0$ to $z \approx 1$ (as
extensively reviewed in Renzini 2006), as well as for other $z>1.4$
ETGs (\cite{cim04,mcc04,daddi05a,sar05,longhetti05,kriek}).

\subsubsection{Extinction and Star Formation Rates}

By giving an age and an $e$-folding time, the best fitting procedure
automatically implies an ongoing SFR for each galaxy. Note however
that the procedure is not specifically optimized to estimate this
quantity, and therefore the results should be regarded only as rough
estimates or upper limits.  Nevertheless, these {\it implied} SFRs are
typically very low, especially in the case of M05 models, as indeed
expected for passive galaxies. For two galaxies (ID 1498 and 2543)
BC03/CB07 models give substantially higher SFRs than for all other
galaxies, and one of them (ID 2543) is the one with the strongest
[OII]$\lambda$3727 flux and the only one marginally detected at
24$\mu$m.

Even assuming the SFRs in Table 3 at face value, the implied specific
star formation rates are very low (i.e. star formation rate per unit
mass, SSFR = SFR/${\cal M}$), as shown in Fig. 10.  These SSFRs are
typical of the lower envelope of the SSFR distribution at $1<z<2$
characteristic of the general population of the oldest galaxies with
just residual, if any star formation (\cite{feulner,juneau,kriek}),
This implies that our sample galaxies cannot increase their stellar
mass significantly at the current rate of star formation. For
instance, they cannot double their mass from the look-back time of
their redshifts to the present day.  The inverse of the SSFR
represents the characteristic secular timescale of the stellar mass
growth.  For our sample galaxies, these timescales are in the range of
30-10$^4$ Gyr. 

The dust extinction inferred from the SED fitting analysis is 
generally very low (with some exceptions with BC03 model fitting). By
forcing $A_V=0$ the results of the SED fitting do not change
substantially: the median age increases by a factor of 2 for BC03 and
$<$25\% for both M05 and CB07, $\tau$ slightly increases by 20\% 
and 5\% for BC03 and CB07 respectively, whereas no significant
changes occur for the stellar masses (5\% for BC03 and $<2$\%
for M05 and CB07). The test with $A_V=0$ indicate that the results of
the SED fitting are stable (especially for M05 and CB07 models)
 and that the major source of uncertainty remains the choice of the 
synthetic spectra library.

\begin{figure}
\centering
\includegraphics[width=\columnwidth]{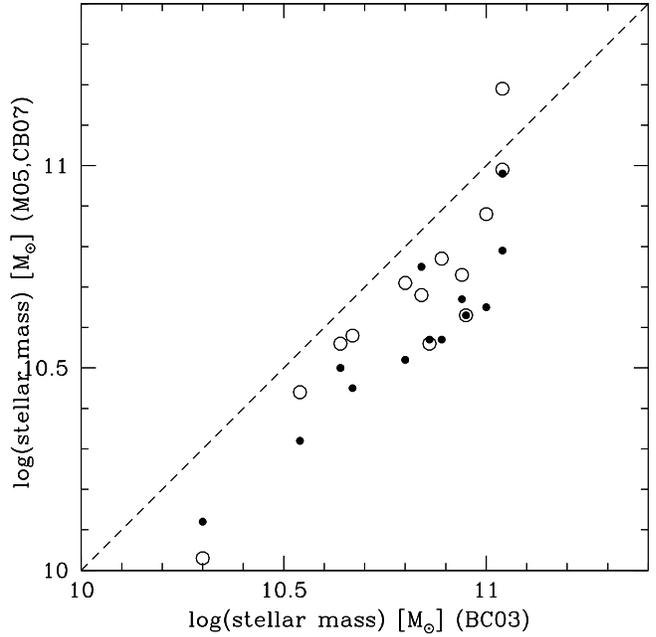}
\caption{Comparison between stellar masses derived with Bruzual \&
Charlot (2003), Maraston (2005) (filled circles) and Charlot \&
Bruzual (2007) (open circles) synthetic spectral templates.}
\end{figure}

The properties inferred with SED fitting are in broad agreement with
the characteristics of the UV spectra, as shown in Fig. 9 where the
stacked spectrum of the passive galaxies is compared with the three
synthetic spectral templates corresponding to the average best fit
results of the photometric SED fitting (see Table 3).

\begin{figure}
\centering
\includegraphics[width=\columnwidth]{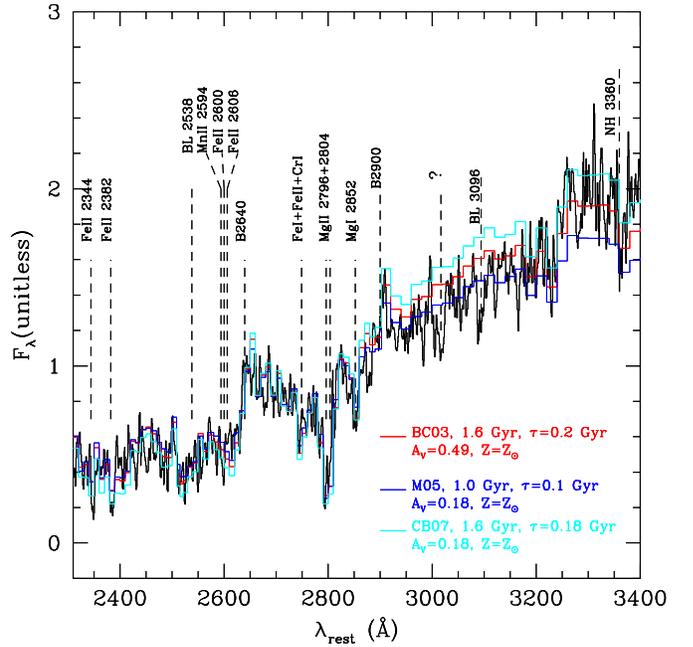}
\caption{Comparison between the average spectrum of GMASS passive
galaxies and the three synthetic spectral templates corresponding to the
average best fit results of the photometric SED fitting. The spectral
resolution of the templates is lower than in the stacked spectrum.}
\end{figure}

\begin{figure}
\centering
\includegraphics[width=\columnwidth]{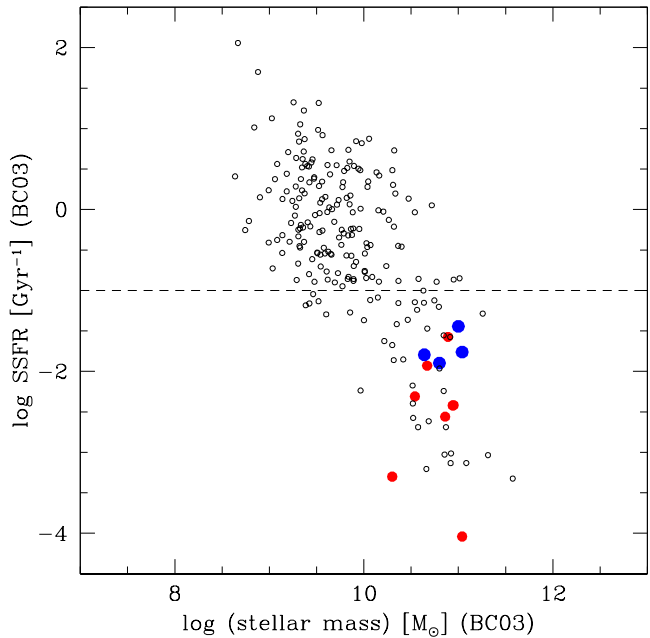}
\caption{The SSFRs of galaxies with star formation activity detected
in the spectra or inferred from the photometric SED fitting. 
The blue circles indicate the four galaxies with observed [OII]$\lambda$3727
emission. The red circles show the galaxies for which the formal SFR
comes from the the photometric SED fitting (with BC03 models). 
With M05 or CB07 models, the SSFRs are even smaller or close to
SSFR$\sim$0. Black circles indicate the general distribution of galaxies
in the same redshift range of our sample ($1.3 \leq z < 2$) using the
data of Feulner et al. (2005). The dashed line indicates the SSFR required to
double a galaxy's mass between the average redshift in this $1.3 \leq
z < 2$ bin and the present (assuming a constant SFR).}
\end{figure}

\begin{figure}
\centering
\includegraphics[width=\columnwidth]{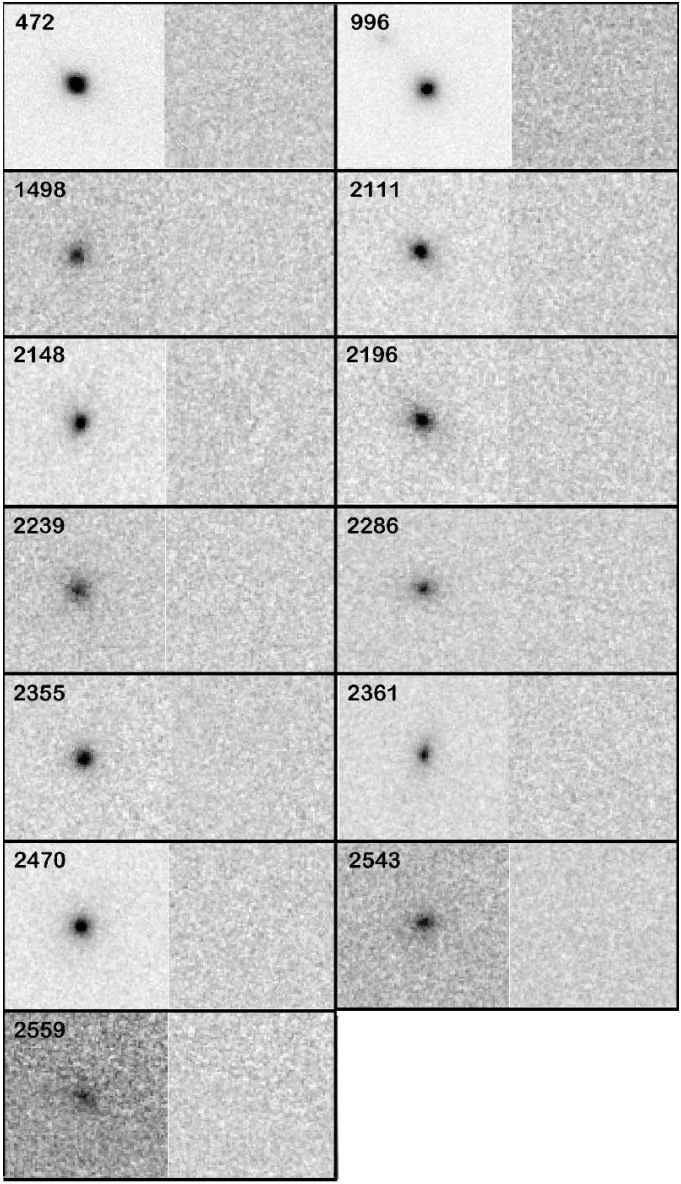}
\caption{Gallery of the 13 passive galaxies in the sample. Each
postage stamp is 2.4$\times$2.4 arcsec$^2$. On the right of each galaxy
cutout, the images of the GALFIT residuals are also shown.}
\end{figure}

\begin{table*}
\caption{Photometric SED fitting results ($Z=Z_{\odot}$)}   % title of
Table
\label{table:1}      % is used to refer this table in the text
\centering                          % used for centering table
\begin{tabular}{l c | c c c c c | c c c c c | c c c c c} % centered columns (4 columns)
\hline\hline                 % inserts double horizontal lines
ID &$M_B$&Age&$A_V$&${\cal M}$&SFR&$\tau$&Age&$A_V$&${\cal M}$&SFR&$\tau$&Age&$A_V$&${\cal M}$&SFR&$\tau$\\
& & BC03 & & & & & M05 & & & & & CB07 & & & \\
\hline                        % inserts single horizontal line
472&-22.09&0.7&0.6&10.67&0.55&0.1&0.8&0.0&10.49&0.14&0.1&0.8&0.3&10.58&0.19& 0.1\\
996&-20.42&1.3&0.6&10.30&0.01&0.1&0.8&0.6&10.16&0.07&0.1&1.3&0.0&10.03&0.01& 0.1\\
1498&-21.88&1.6&0.6&10.89&2.07&0.3&0.7&0.4&10.61&0.45&0.1&2.0&0.0&10.77&0.43& 0.3\\
2111&-21.80&2.3&0.1&10.86&0.20&0.3&1.1&0.0&10.61&0.01&0.1&1.1&0.1&10.56&0.01& 0.1\\
2148&-22.52&1.3&0.4&11.04&0.01&0.1&1.3&0.2&11.02&0.01&0.1&2.8&0.0&11.19&0.09& 0.3\\
2196&-22.15&1.3&0.3&10.84&0.00&0.1&1.3&0.0&10.79&0.00&0.1&1.1&0.1&10.68&0.01& 0.1\\
2239&-21.40&1.3&0.5&10.64&0.00&0.1&1.1&0.2&10.54&0.01&0.1&1.0&0.5&10.56&0.02& 0.1\\
2286&-21.39&2.0&0.5&10.80&0.47&0.3&1.1&0.2&10.56&0.01&0.1&2.3&0.1&10.71&0.14&0.3\\
2355&-21.62&0.8&0.6&10.54&0.17&0.1&0.9&0&10.36&0.04&0.1&0.8&0.4&10.44&0.14&0.1\\
2361&-21.85&2.8&0.2&11.04&0.07&0.3&1.4&0.2&10.83&0.00&0.1&3.0&0.0&10.99&0.03& 0.3\\
2470&-22.07&2.2&0.1&10.94&0.33&0.3&1.1&0.0&10.71&0.01&0.1&1.0&0.3&10.73&0.03& 0.1\\
2543&-21.30&1.1&1.8&11.00&12.4&0.3&1.0&0.6&10.69&0.03&0.1&2.1&0.6&10.88&0.40& 0.3\\
2559&-22.09&2.2&0.1&10.95&0.34&0.3&1.0&0.0&10.67&0.03&0.1&1.1&0.0&10.63&0.01& 0.1\\
&&&&&&&&&&&&&&&&\\
Mean&-21.79&1.6&0.4&10.81&0.35&0.2&1.1&0.2&10.62&0.06&0.1&1.6&0.2&10.67&0.12& 0.2\\
$rms$&0.53&$\pm$0.6&$\pm$0.2&$\pm$0.22&$\pm$0.57&$\pm$0.1&$\pm$0.2&$\pm$0.2&$\pm$0.28&$\pm$0.12&$\pm$0.0&$\pm$0.8&$\pm$0.2&$\pm$0.28&$\pm$0.15& $\pm$0.1\\
\hline                                   %inserts single line
\end{tabular}
\small
Rest-frame absolute magnitudes ($M_B$) are relative to BC03 fitting.
The three groups of columns (Age, $A_V$, {\cal M}, SFR, $\tau$) are
relative to BC03, M05 and CB07 models respectively. The mean and rms of
$A_V$ and SFR for BC03 models are given excluding the highly discrepant
values for object 2543. The units are Gyr, mag,
log$({\cal M}$/M$_{\odot}$), M$_{\odot}$yr$^{-1}$ and Gyr for Age, 
$A_V$, Mass, SFR, $\tau$ respectively.

\normalsize
\end{table*}

\section{Morphological properties}

\subsection{The HST Data}

Thanks to the availability of deep and ultradeep HST imaging from
GOODS and HUDF projects, it was possible to complement the spectroscopy
and photometry with quantitative morphological information. 
GOODS has surveyed the Chandra Deep Field South (CDFS) with ACS and
filters F435W (B), F606W (V), F775W (I) and F850LP (z), with exposure 
times respectively of 3, 2.5, 2.5 and 5 orbits. HUDF has instead surveyed 
a smaller area of about $3\times3$ arcmin$^2$ which has a large overlap
with the GMASS field (see Kurk~et~al.~2007a for further details) in
the same 4 bands, but with exposure times respectively of 56, 56, 150
and 150 orbits. For both datasets the publicly available images were 
drizzled to a scale of 0.03 arcsec/pixel in order to improve the PSF
pixel sampling. Furthermore, the HUDF was observed also with deep NICMOS 
imaging of $2.5\times2.5$ arcmin$^2$ with F110W and F160W filters 
(\cite{thompson}). 
NICMOS images were drizzled to a final scale of 0.09 arcsec/pixel.

\subsection{Visual classification}

A visual classification based on eye inspection was performed
on the whole GMASS sample independently by two of us (PC and GR). 
The global morphological results will be presented in a forthcoming
paper (Cassata et al. 2007, in preparation). The visual analysis was always
done in the ACS filter band closest to the rest-frame $B$-band. 
For the objects at $z>1$ (like the ones presented in this
paper), the analysis was done in the reddest filter available (z-band). 
The classification scheme is based on 6 classes (see also Cassata
et al. 2005): 1. Ellipticals/S0 galaxies; 2. Peculiar ETGs 
(i.e. ETGs with signs of interaction or with some isophotal
asymmetry); 3. Normal spirals; 4. Perturbed spirals, that is disk
galaxies with an evident bulge but showing also asymmetries due to
interactions or regions of enhanced star formation; 5. Irregular
galaxies; 6. Compact objects. In Figure 11 we present a gallery of 
the 13 galaxies in the sample. Each box is 2.4$\times$2.4 arcsec$^2$, 
corresponding to about 20 kpc at redshift 1.5. 

Fig. 11 clearly shows that the majority of passive galaxies have 
the spheroidal morphology characteristic of ETGs. In particular,
we classified 7 of them as pure ellipticals, as they appear resolved, 
concentrated and very regular; 2 of them have been classified as 
spirals (2239 and 2559), as they are not very concentrated; finally, 
4 have been placed in the compact class, i.e. with very concentrated and 
regular morphological structure (472, 2286, 2361 and 2543) (see Table 4). 
The NICMOS + F160W images available in the HUDF for the galaxies IDs 
472 and 996 show very compact and spheroidal
morphologies at $\lambda_{rest} \approx 0.5-0.6 \mu$m (Fig. 12).

\subsection{Surface brightness profile analysis}

We used GALFIT (Peng~et~al.~2002) to model the surface brightness
distribution for the galaxies in our sample. As these galaxies are 
very red and become rapidly very faint going to short wavelengths, 
GALFIT was run on the image taken with the reddest available ACS 
filter (z-band, filter F850LP). The PSF with which GALFIT
convolves the models during the fit process was obtained by averaging
10 stars in the field. Two of the objects (472 and 996) lie in HUDF
area, so they have been analyzed in the HUDF ACS F850LP image, while 
the other 11 have been analyzed in GOODS image. We also attempted to
analyze galaxies IDs 472 and 996 in the HUDF NICMOS + F160W image, 
but the fitting did not provide stable and reliable results due to
their small sizes with respect to the pixel scale and the 
difficulty to derive the PSF because of the very few suitable 
stars available. 

In all cases, we used a S\'ersic profile to model the galaxy 
surface brightness profiles :

\begin{equation}
I(r)=I_eexp\left\{-b_n\left[\left(\frac{r}{r_e}\right)^{1/n}-1\right]\right\}
\end{equation}

where $I(r)$ is the surface brightness measured at distance $r$, $I_e$
is the surface brightness measured at the effective radius $r_e$ and
$b_n$ is a parameter related to the sersic index $n$. For $n$=1 and $n$=4
the Sersic profile reduces respectively to an exponential or
deVaucouleurs profile. Bulge dominated objects having typically high 
$n$ values (e.g. $n>2$) and disk dominated objects having $n$ around unity.

\begin{figure*}
\includegraphics[width=4cm]{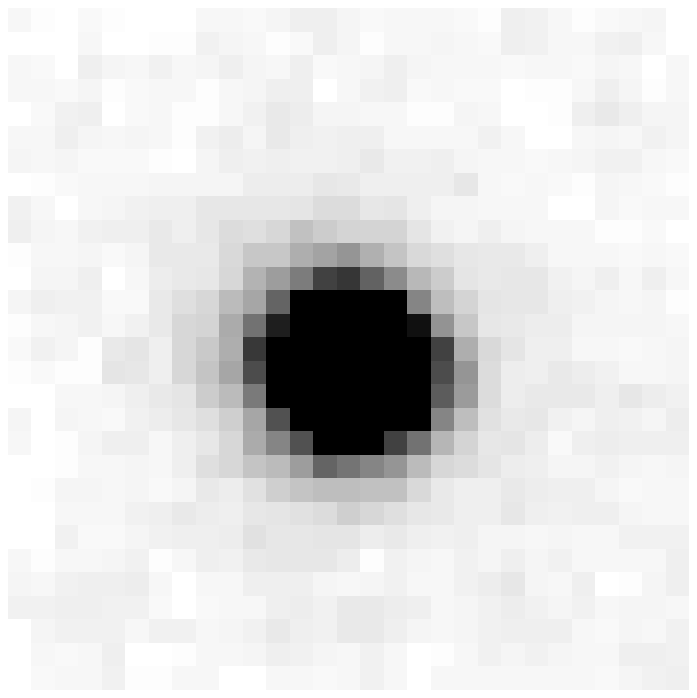}
\includegraphics[width=4cm]{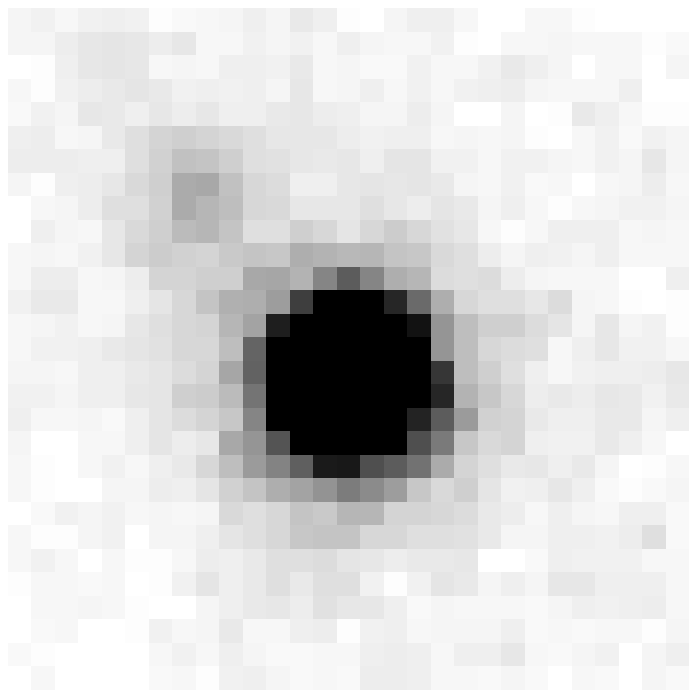}
\caption{NICMOS images (2.4$\times$2.4 arcsec$^2$) of IDs 472 and 996.}
\end{figure*}

\subsection{Simulations}

To test the capability of GALFIT to obtain reliable size measurements we
made use of simulations. To this purpose, we built a set of galaxies
with a wide range of properties, to take into account all possible
systematic biases. In particular, we have chosen a grid of values for
magnitudes, radii, Sersic indices and axial ratios. Apparent magnitudes 
$m_z$ span from 22.5 to 24.5, reflecting the magnitude distribution of the
13 real galaxies in the sample. The range of radii goes from 0.06 to 
0.45 arcsec (basically from 2 pixels to the typical size that local galaxies 
with masses comparable to those of our sample would have at $z\sim
1.4-2$). Sersic indices $n$ span 
from 1 to 6 and axial ratios $b/a$ varied between 0.4 and 1. A set of
1000 galaxies was built by randomly selecting the above parameters 
from their intervals. Then, the galaxies were simulated using GALFIT,
and convolved with the same average star PSF used for the fitting process.
Finally, the simulated galaxies were placed in the real z-band 
ACS/GOODS-South image in random positions in order to take into account 
possible systematics due to deblending and variations of the local 
background. The same GALFIT fitting process used for the 13 real galaxies 
was then applied to derive the morphological parameters of the simulated
galaxies. First, we used \emph{SExtractor} to obtain first estimates for 
magnitude, size, axial ratio and orientation. Second, we ran GALFIT using 
the \emph{SExtractor} values as first guesses. In the fitting procedure, 
$n_{Sersic}$ was left free to vary between 0.5 and 8. The noise images 
requested by GALFIT to estimate $\chi^2$, were built using proper values 
of RMS, GAIN end EXPTIME.

\begin{figure}
\begin{center}
\includegraphics[width=\columnwidth]{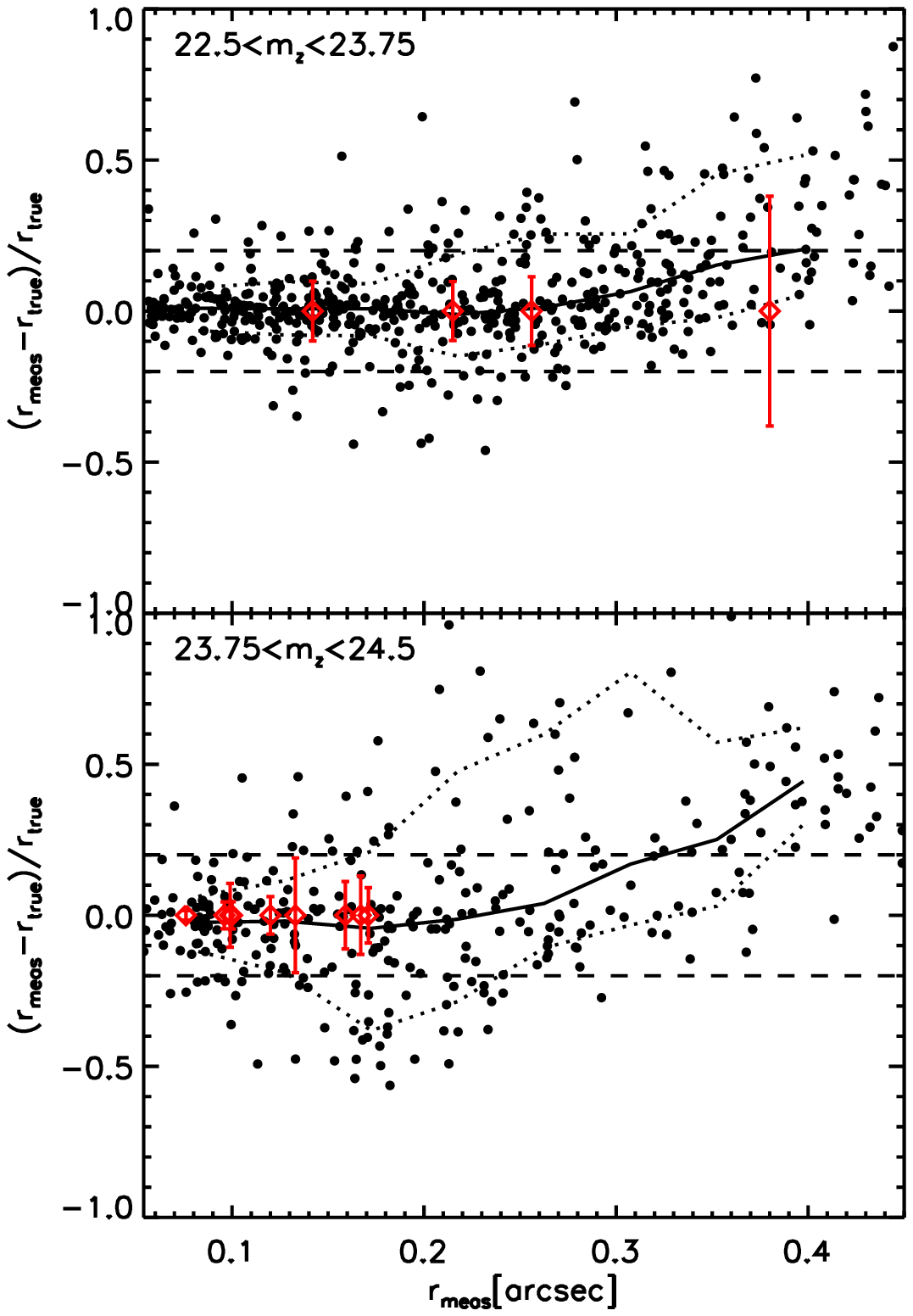}
\end{center}
\caption{The fractional difference between measured (r$_{\rm meas}$) and input 
(r$_{\rm true}$) sizes for the 1000 simulated galaxies, as a function 
of the output size. The upper and the lower panel show the results 
for galaxies brighter and fainter than $m_z$=23.75 respectively. The solid
line shows the median of the error distribution, and the dotted lines 
includes 68\% of the objects. The dashed lines indicate the limiting regions
relative to errors of $\pm$20\%. Red symbols show the measured sizes
for the 13 galaxies in the sample, together with the formal error
measured by GALFIT.}
\label{simu} 
\end{figure}

The results of this process are summarized in Fig.~\ref{simu}, where
we show the fractional error on the measured size as a function of the output
size. This allows a simple comparison between the sizes and errors 
estimated with this technique for our 13 real galaxies and the results 
of simulations. It can be noted that for small sizes the average error 
is very close to zero, with a small scatter. For large sizes, instead, 
the error distribution moves towards positive values, and the scatter 
increases. This tendency to overestimate sizes for large objects is 
also dependent on the magnitude, being mild for galaxies with $m_z<23.75$ 
and more important for object with $m_z>23.75$. In any case, this 
overestimate is relevant only for objects with sizes larger than $0.25-0.3$ 
arcsecs. In the same Figure, we also show the radii measured for the 
13 galaxies in the sample, together with the formal errors determined by 
GALFIT. It can be seen that 4 and 9 of them are respectively brighter and 
fainter than $m=23.75$, and that 12 have measured sizes smaller than 
$\sim0.25$ arcsec, in the regime where the simulations show no systematic 
errors and a small scatter. Only one object has a measured size larger 
than 0.3 arcsec, for which the simulation shows a typical overestimate of 
the size of about 20\%. The size for this object can thus be considered 
as an upper limit.

Overall, the errors measured by GALFIT are smaller than the scatter
shown by simulated galaxies with similar output sizes.  Most of the 13
galaxies have sizes for which the simulation shows a scatter of about
20\%, so we decided to assign a minimum fiducial error of 20\% to our
galaxy sizes.

\subsection{GALFIT results}

Once assessed the capability of GALFIT of retrieving robust
measurements of the size even for galaxies at high redshift, we run
the code on our 13 galaxies, using \emph{SExtractor} extimates of
magnitude, size, axial ratio and position angle as first guesses. 
The GALFIT results are shown in Table 6. The residual maps provided 
by GALFIT, as result of the subtraction of the best fit model from 
the original galaxies, show no significant structures for all the 
galaxies (Fig. 11). 11 of the 13 objects have best fit solutions with 
Sersic index $n>2$, indicating that the bulk of the galaxy light comes 
from a dominant bulge component (e.g. in Fig. 14). 
In our sample, among the 7 galaxies 
visually classified as pure ellipticals, 6 have $n\sim$4. In the class
of visually classified ellipticals, only galaxies 1498, 2355 and 
2559 are best fitted by a smaller value of the Sersic index ($n\sim$1-2).
The GALFIT results broadly agree with 
what was anticipated by the visual classification, i.e. that passive 
galaxies at $z\sim$1.4--2 have the typical morphology of ETGs, as in 
the present--day Universe.

\begin{figure}
\centering
\includegraphics[width=\columnwidth]{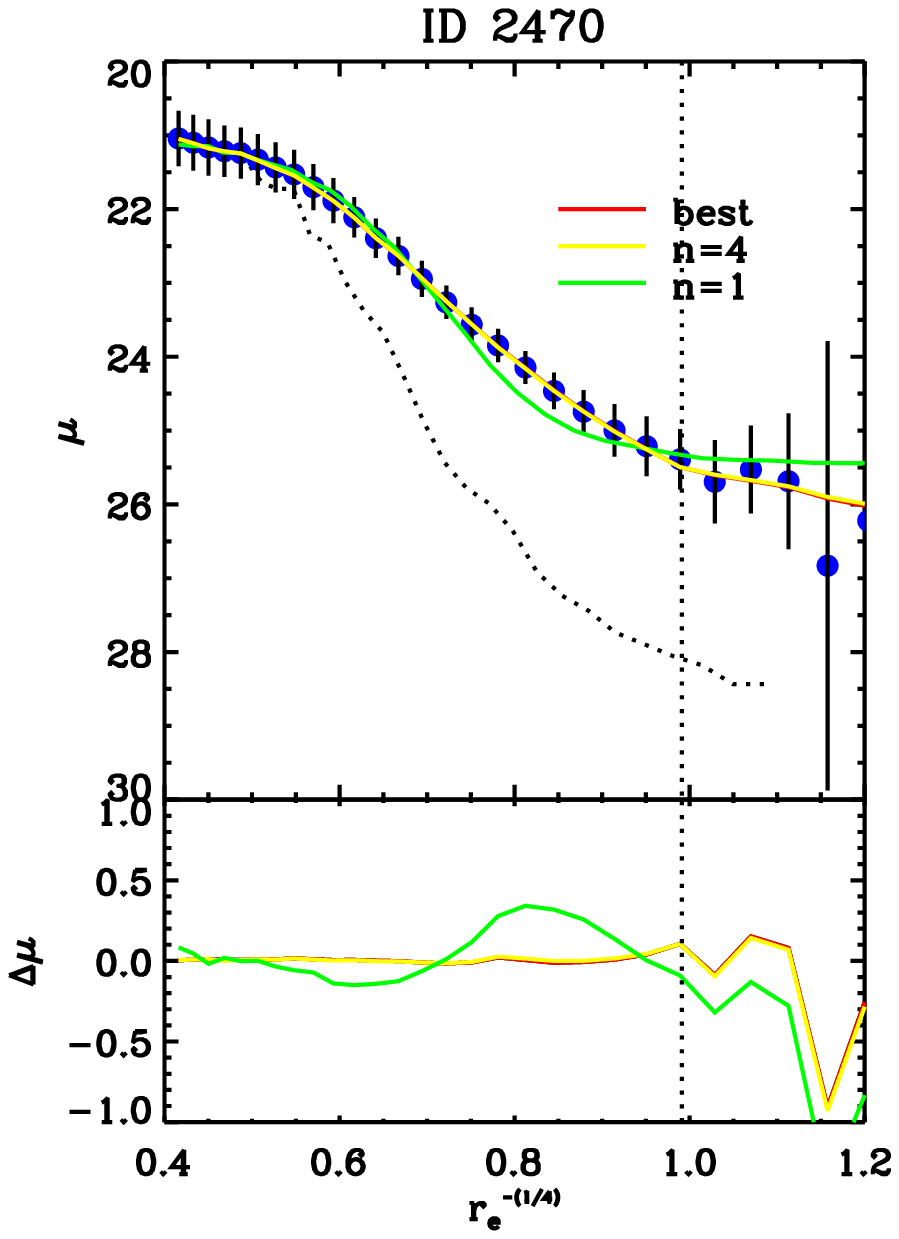}
\caption{Example of an ACS z-band surface brightness profile fit 
expressed in mag arcsec$^{-2}$ ($\mu$) relative to 
the galaxy ID 2470 where a $r^{1/4}$ profile provides a very good 
match with the data (with $r$ expressed in arcsec). The black dotted, 
red, yellow and green lines 
indicate respectively the PSF profile, the formal best fit profile, 
the profile for $n=4$ (de Vaucouleurs) and the profile for $n=1$ 
(exponential). The bottom panel shows the residuals between the best
fit curves and the observed profile.}
\end{figure}

The galaxy angular sizes were circularized as $r_e = a_e(1-\epsilon)^{1/2}$,
where $a_e$ and $\epsilon$ are the effective radius along the major axis 
and the ellipticity of the galaxy morphology respectively. The values are in
the range of 0.07--0.38 arcsec, corresponding to 0.6--3.2 kpc (Table 4).  
Of the two galaxies located in the HUDF (IDs 472 and 996) previously 
identified by Daddi et al. (2005a) (IDs are 3650 and 8238 respectively),
ID 472 has $n$ and $r_e$ consistent within the errors with those
of Daddi et al. (2005), whereas for ID 996 we measure a significantly 
smaller $r_e$. Here we adopt our measurements for ID 996.

In our morphological analysis, the reddest filter available for all 
galaxies (ACS F850LP), does not cover the rest-frame optical, but 
the rest-frame mid--UV ($\approx$3000--4000 \AA, i.e. approximately 
the $U$-band). This might introduce a bias due to the possible
morphological $K$-correction and/or the internal color gradients usually 
present in ETGs, thus resulting in a wavelength-dependent size of the 
galaxies. In particular, gradients with colors redder towards the
center (as often observed in ETGs; e.g. Peletier et al. 1990) could 
make the sizes measured in the optical larger than those in the UV. 
However, McIntosh et al. (2005) showed that 
for this kind of gradient, the expected size variation from the 
rest-frame $U$-band to the $R$-band is $\Delta r / r_{obs} \approx 
-0.075(z-0.47)$. This means that our galaxies would 
have sizes larger by only $\approx$6--11\% if observed in the 
rest-frame $R$-band for the redshift range of our sample ($1.4<z<2$). 
Other results indicate that the sizes of spheroidal galaxies do not 
change substantially as a function of wavelength (e.g. by comparing 
$r_e$ measured in ACS and NICMOS bands; e.g. see \cite{mcg1,trujillo07}).

As discussed in the next sections and found in other works (e.g. 
\cite{daddi05a,trujillo06}), the sizes derived for most of our sample 
galaxies are much smaller than the ones observed at $z \sim 0$ in 
ETGs with the same stellar mass. We explored the 
possibility that these small $r_e$ may be partly due to the
presence of a point-like unresolved central source which dominate 
the surface brightness profiles and bias the estimate of $r_e$ 
towards small values. Thus, we attempted to perform fits with 
a two-component surface brightness profile made by a central
unresolved source (using the stellar PSF) plus a Sersic profile. 
In all cases we found no significant improvements in the resulting 
best fit profile with respect to those obtained with a single (Sersic)
component. More importantly, we found that typically less than 16\% of 
the light (corresponding to 
$\Delta$mag$>$2) is located in the putative central component, and
the estimated values of $r_e$ are all consistent with the ones
obtained with the single component fitting within 10\%. This 
experiment strengthens even more that the $r_e$ measured
for most galaxies in our sample are reliable and indeed very small.

\begin{table}
\caption{Surface brightness fitting results (ACS F850LP band)}   % title of Table
\label{table:1}      % is used to refer this table in the text
\centering                          % used for centering table
\begin{tabular}{c c c c c c}        % centered columns (4 columns)
\hline\hline                 % inserts double horizontal lines
ID & Class & n & $r_e$ & $R_e$ & $\mu_e$ \\     
& & & arcsec & kpc & mag arcsec$^{-2}$ \\     
\hline                        % inserts single horizontal line
472&6&4.1$\pm$0.7&0.076&0.64$\pm$0.13&15.52$\pm$0.43\\
996&1&4.6$\pm$0.7&0.100&0.84$\pm$0.17&17.78$\pm$0.43\\
1498&1&1.5$\pm$0.2&0.120&1.01$\pm$0.20&16.71$\pm$0.43\\
2111&1&4.0$\pm$0.4&0.099&0.84$\pm$0.17&16.39$\pm$0.43\\
2148&1&3.7$\pm$0.3&0.142&1.20$\pm$0.24&16.45$\pm$0.43\\
2196&1&6.0$\pm$0.9&0.380&3.22$\pm$1.22&18.95$\pm$0.82\\
2239&4&2.2$\pm$0.2&0.256&2.16$\pm$0.43&18.84$\pm$0.43\\
2286&6&2.6$\pm$0.3&0.159&1.35$\pm$0.27&17.82$\pm$0.43\\
2355&1&2.2$\pm$0.2&0.096&0.81$\pm$0.16&16.48$\pm$0.43\\
2361&6&4.1$\pm$0.4&0.133&1.13$\pm$0.23&16.97$\pm$0.43\\
2470&1&4.2$\pm$0.3&0.215&1.81$\pm$0.36&17.78$\pm$0.43\\
2543&6&2.2$\pm$0.3&0.167&1.41$\pm$0.28&18.02$\pm$0.43\\
2559&4&1.0$\pm$0.5&0.171&1.43$\pm$0.29&17.25$\pm$0.43\\
\hline                                   %inserts single line
\end{tabular}
Class : visual classification (Section 6.2)
\end{table}

\section{The size -- mass relation}

Early-type galaxies at $0<z<1$ follow a well-defined size--mass 
relation (e.g. Shen et al.  2003; McIntosh et al. 2005), with
the size increasing as a function of mass. Models of galaxy
formation predict different size-mass relations depending on
the history of mass assembly (see Shen et al. 2003). 

The role of GMASS is the secure spectroscopic identification of a 
significant number of passive galaxies down to lower masses than 
in previous samples at $1.4<z<2$ which were biased towards the most
massive systems (e.g. \cite{daddi05a,trujillo06,mcg1,trujillo07}).
This allows us to better probe the size--mass relation of ETGs at
$z\approx$1.4--2 down to $\approx 10^{10}$ M$_{\odot}$.

\begin{figure*}
\begin{center}
\includegraphics[width=\columnwidth]{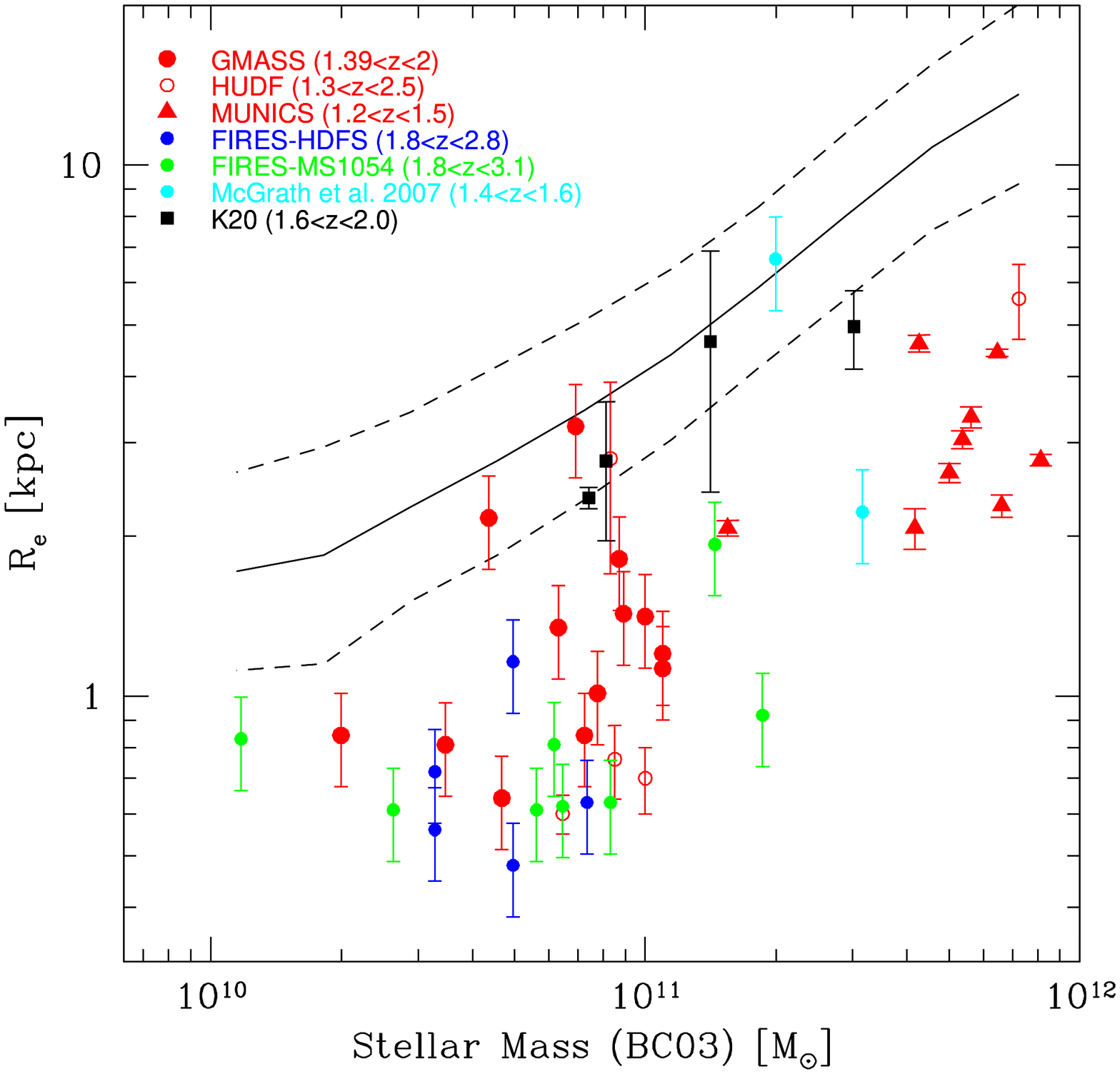}
\includegraphics[width=\columnwidth]{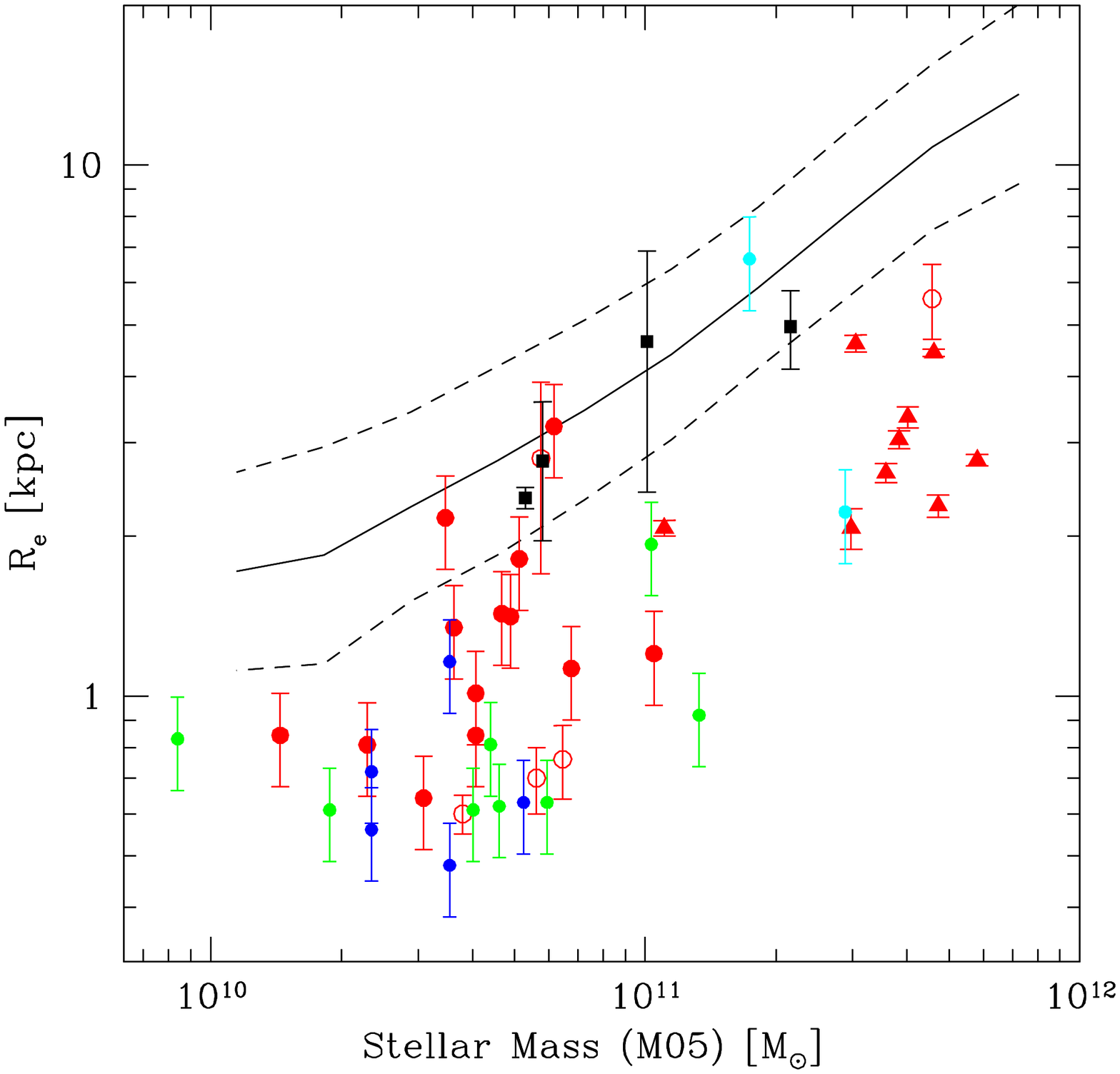}
\end{center}
\caption{The distribution of physical sizes versus stellar mass for
passive galaxies in the GMASS sample (red filled circles).
For clarity, error bars are not shown. Stellar masses are estimated 
with BC03 (left panel) and M05 (right panel) spectral synthesis models. 
The solid line shows the local size--mass relation of ETGs as derived with 
the SDSS sample by Shen~et~al.~2003, with the dashed lines indicating the 
$\pm 1 \sigma$ scatter around this relation. Blue, green, cyan
filled circles indicate the passive galaxies of Zirm~et~al.~2007
(FIRES - HDFS), Toft et al. 2007 (FIRES - MS1054) and McGrath et al. 
2007b respectively, the red triangles show the MUNICS massive galaxies 
of Trujillo~et~al.~2006 and Longhetti et al. 2007, the red 
open circles show the HUDF passive galaxies of Daddi~et~al.~2005 and 
Maraston et al. (2006), the black filled squares the K20 galaxies of 
Cimatti et al. (2004). The two galaxies of McGrath et al. 2007b whose
SEDs were fitted with CB07 templates are located in the right panel.}
\label{mass-size} 
\end{figure*}

The size -- stellar mass relation for the passive galaxies in our 
sample is presented in Figure 15 and compared with the local 
size-mass relation measured by Shen~et~al.~2003 for ETGs 
in the SDSS and with other passive galaxies at $z>1.2$ from the
literature. We note that in the local relation published by Shen et al. (2003, 
their Table 1) there was a typo in the $b$ parameter, 
while here we used the correct one ($b=2.88 \times 10^{-6}$; S. 
Shen, private communication).

In the comparison amongst different samples, it is important to 
account for the differences in the adopted synthetic stellar population 
models and IMFs. Throughout the present work, we adopt a Chabrier IMF 
and scale all the literature masses to that IMF with the following relations: 
log${\cal M}$(Chabrier)=log${\cal M}$(Salpeter)--0.23 and log${\cal
M}$(Chabrier)=log${\cal M}$(Kroupa)--0.04. In addition to the IMF conversion,
it is sometimes necessary to scale stellar masses by taking into
account the different model spectra used to fit the SEDs. 
Based on previous (e.g. \cite{wuyts07,
maraston06}) and present results (Section 5.2), we adopted the conversion 
log${\cal M}$(M05)=log${\cal M}$(BC03)--0.15 in order to take into
account the average decrease ($\approx$40--50\%) of M05 masses with 
respect to BC03 results for galaxies at $z\approx 1-3$.

Shen et al. (2003) adopted a Kroupa IMF and used the stellar masses 
that were previously estimated by Kauffmann et al. (2003) through the 
comparison of observed features in the SDSS spectra with BC03 template 
spectra. However, we decided not to scale the local size--mass relation 
of Shen et al. (2003) to account for the difference between BC03 and M05 
masses because at $z\sim0$ the typical ages of ETGs 
are so old (several Gyr) that the effects of TP-AGB stars should be
negligible. For the sample of Daddi et al. (2005a) we used the most recent 
SED fitting results of Maraston et al. (2006) who adopted a Kroupa IMF and 
provided stellar masses estimated with both BC03 and M05 models. 
Trujillo et al. (2006) adopted a Kroupa IMF and their stellar masses refer
to BC03 models. For the sizes of the MUNICS galaxies, we 
adopted those of Longhetti et al. (2007). For the FIRES sample galaxies, 
Zirm et al. (2007) used M05 models and a Salpeter (1955) IMF, whereas Toft
et al. (2007) 
used BC03 models and a Salpeter IMF. We recall that the redshifts of 
the FIRES galaxies of Zirm et al. (2007) and Toft et al. (2007)
are photometric. We also added two spheroidal galaxies at $z \sim 1.5$ 
selected from the sample of McGrath et al. (2007a, 2007b) who adopted 
BC03 (with Chabrier IMF) for their SED fitting.

Fig. 15 shows that passive galaxies at $z>1.2$ follow a clear size--mass 
relation. However, the majority of them has sizes significantly smaller 
than at $z\approx$0 for a fixed stellar mass, whereas only 15-20\% 
are located within the scatter region of the $z\sim0$ relation of 
Shen et al. (2003). 

\begin{figure*}
\begin{center}
\includegraphics[width=\columnwidth]{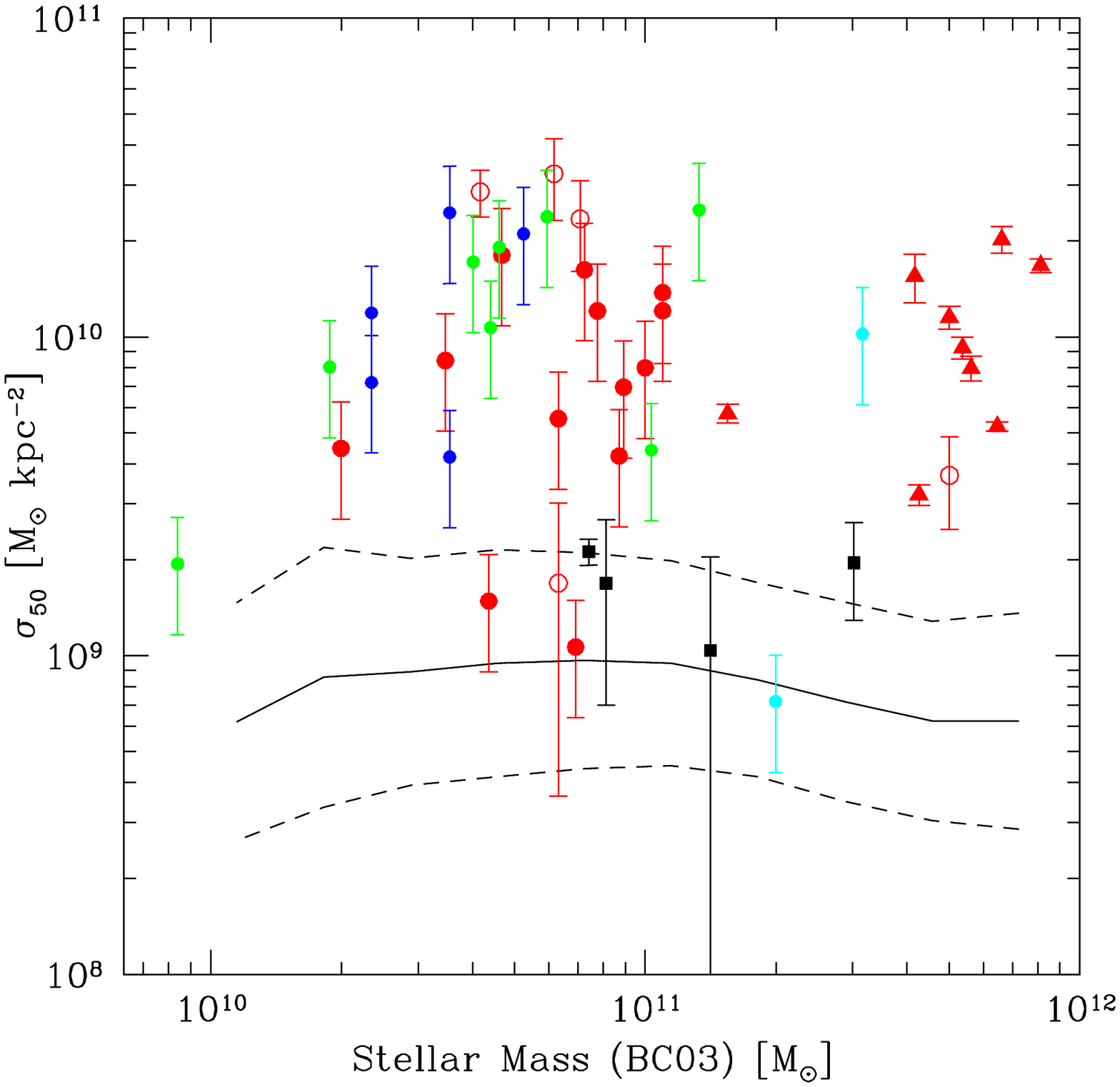}
\includegraphics[width=\columnwidth]{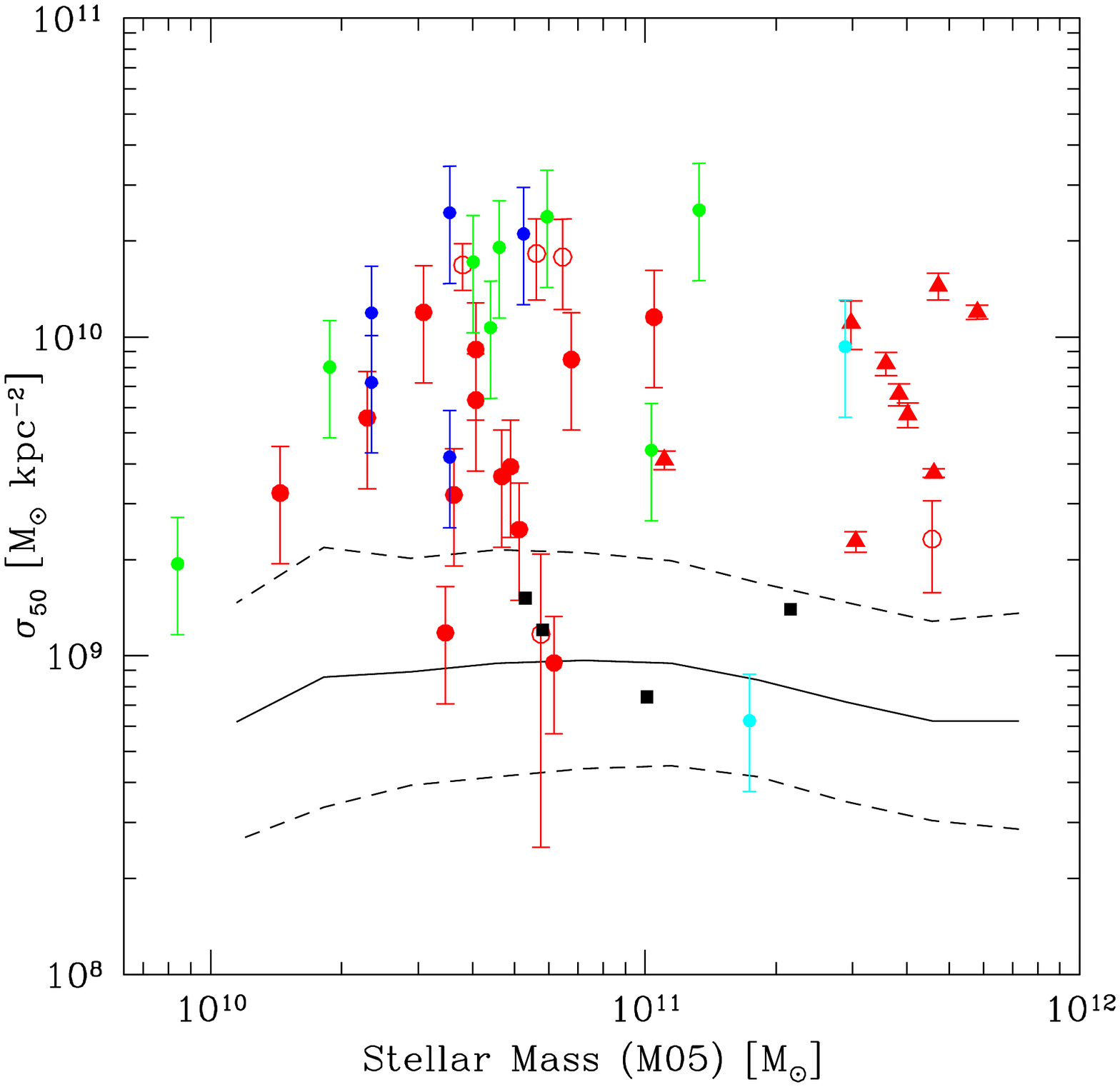}
\end{center}
\caption{
The distribution of stellar mass surface density ($\sigma_{50}$) versus 
stellar mass for passive galaxies. The symbols are the same of Fig. 15.  
The solid line shows the local size--mass relation from the SDSS sample 
by Shen~et~al.~2003, with the dashed lines indicating the scatter around 
this relation.}
\label{mass-size} 
\end{figure*}

The result on the small
sizes at $z>1.2$ holds independently of the adopted stellar masses 
(cf. Fig 15 left and right panels). For the GMASS galaxies 
only (for which we also have stellar 
masses estimated with CB07 models), the results are very similar to those 
of M05 stellar masses if CB07 stellar masses are used. The average ratio 
between the observed sizes and the size at $z\sim0$ (Shen et al. 2003) 
for the same stellar mass (all scaled to BC03 + Chabrier IMF) is 
$\langle R_e(z)/R_e(0) \rangle$= 0.42$\pm$0.25 and 
0.29$\pm$0.14 for $1<z<2$ ($\langle z \rangle$=1.60) and $2<z<4$ ($\langle
z \rangle $=2.52) respectively. 
No significant correlation was found between the galaxy size $R_e$ 
(or $\langle R_e(z)/ R_e(0) \rangle$) 
and physical or environmental properties (age of the stellar population,
SFR, SSFR, rest-frame colors, $A_V$, location inside our outside the 
structure at $z=1.61$). 

The sizes smaller by a factor of $\approx$2-3 than at $z\sim0$ imply that the
stellar mass surface densities are $\approx$5--10 times larger. Fig. 16 
shows the stellar mass surface densities ($\sigma_{50}$) of the passive 
galaxies as a function of stellar mass and the comparison with the
$z\approx0$ relation derived from Shen et al. (2003) sample. The stellar 
mass surface densities are estimated as $\sigma_{50}= 0.5{\cal M} / \pi 
R_e^2$. The difference with respect to $z\sim 0$ galaxies becomes 
even more striking if stellar volume densities are considered: an excess 
by a factor of 3 in size ($R_e$) corresponds to a factor of $\approx$30 
in volume density. 
	
\section{Comparison with early-type galaxies at $z\approx1$}

We also explored how the picture changes at lower redshifts
using the small, but highly complete spectroscopic sample of K20 Survey 
(\cite{cim02}) ETGs at $0.8<z<1.3$ for which also the absorption line 
velocity dispersion
($\sigma_V$) is available. Fig. 17 shows that at $z\sim$1 more galaxies 
can be located within the $z\sim0$ relations (see also \cite{mci,trujillo07}). 

From the scaling relation of ETGs (e.g. \cite{jorgen,vds03}), it
is expected that the densest systems should have necessarily
the highest velocity dispersion. Fig. 17 shows that while ETGs with 
high $\sigma_V$ are located both inside and outside the $R_e$--mass 
relation at $z\approx0$, most of the smallest/densest outliers 
have high values of $\sigma_V$.
The migration of passive galaxies in the $R_e$--${\cal M}$ and
$\sigma_{50}$--${\cal M}$ planes from regions of small size and
high density to the $z \approx 0$ relations (Fig. 15--17) is further 
evidence that the redshift range of about $1<z<2$ is the critical
cosmic epoch for the assembly and structural transformation of
ETGs (e.g. \cite{fon04,glaze,abraham07,arnouts}).

\begin{figure*}
\begin{center}
\includegraphics[width=\columnwidth]{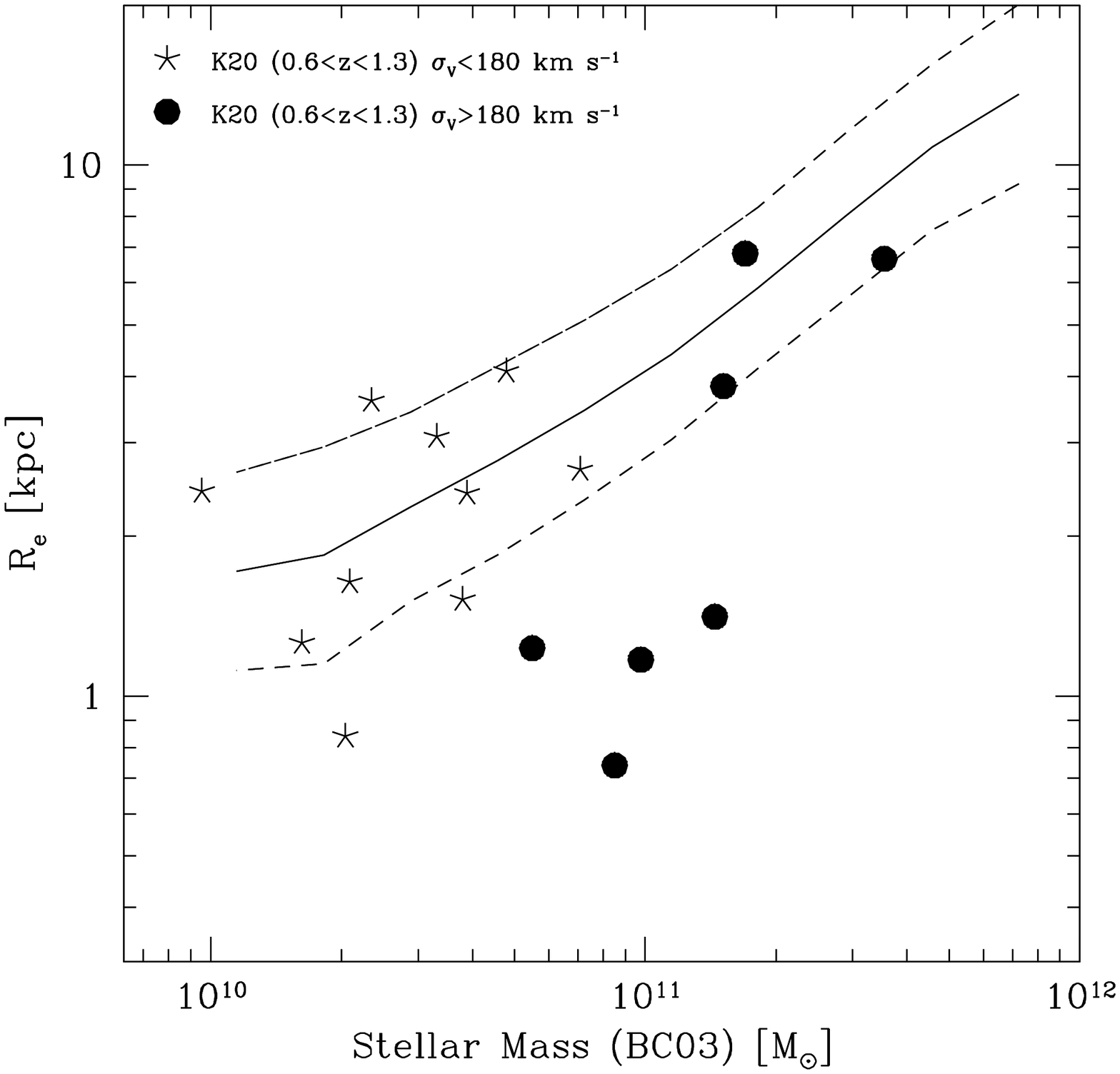}
\includegraphics[width=\columnwidth]{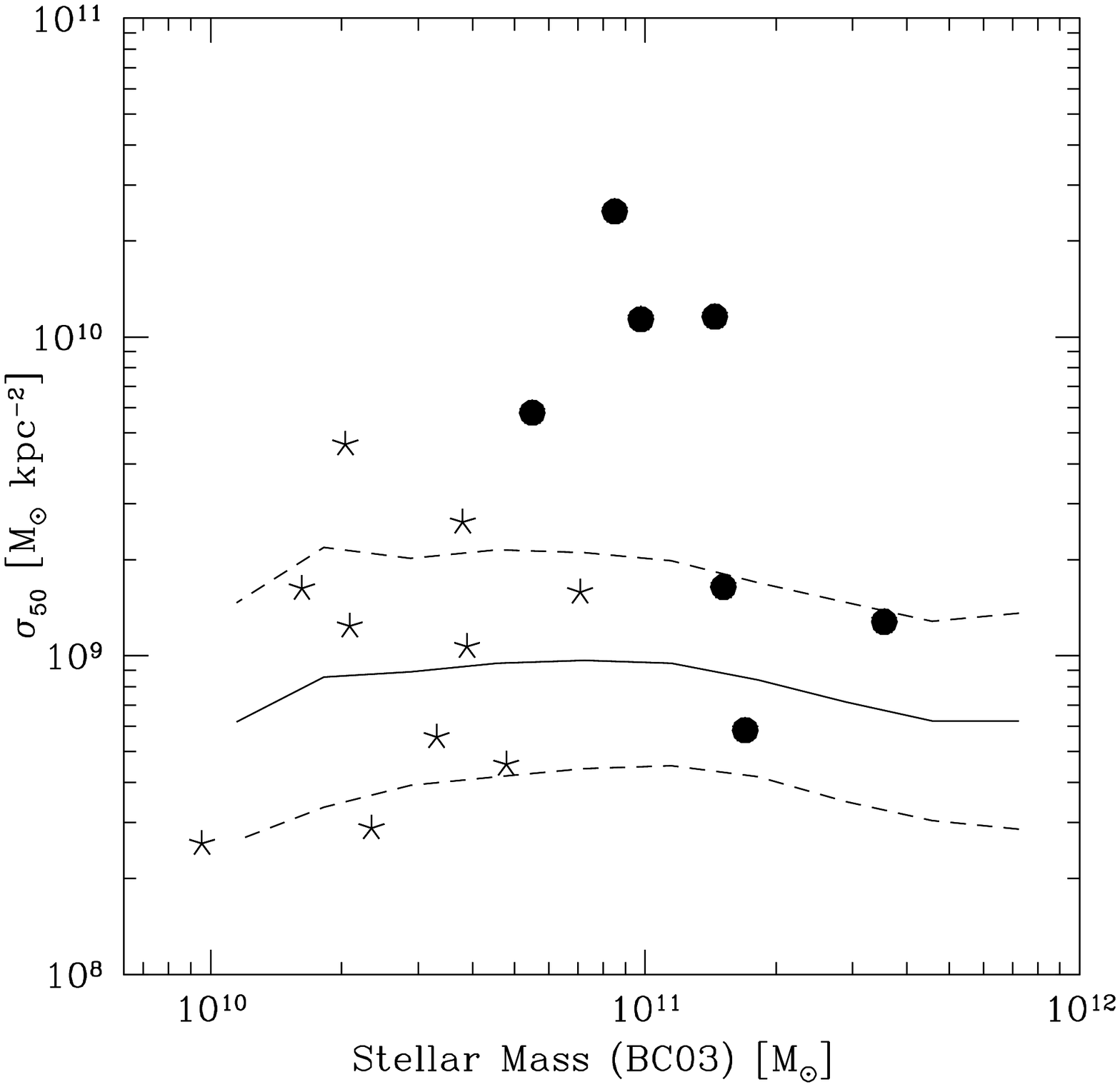}
\end{center}
\caption{{\it Left panel:} the distribution of physical sizes 
versus stellar mass for
ETGs at $z\approx1$. The solid line shows the local 
size--mass relation of ETGs as derived with the SDSS sample 
by Shen~et~al.~2003, with the dashed lines indicating the scatter around 
this relation. The starred and filled symbols indicate the K20 ETGs 
from di Serego Alighieri et al. (2005) with velocity dispersion
$\sigma_V<$180 km s$^{-1}$ and $\sigma_V>$180 km s$^{-1}$ respectively.
{\it Right panel:} the distribution of stellar mass surface density 
($\sigma_{50}$). The symbols are the same of left panel.}
\label{mass-size} 
\end{figure*}

\section{The Kormendy relation}

In the absence of the velocity dispersion measurements needed to study
where the GMASS passive galaxies lie in the Fundamental Plane of ETGs,
we explored how they behave with respect to the Kormendy relation, i.e.
a projection of the Fundamental Plane which correlates $\mu$, the mean
surface brightness within $r_e$ and the physical size $R_e$.

\begin{figure}
\begin{center}
\includegraphics[width=\columnwidth]{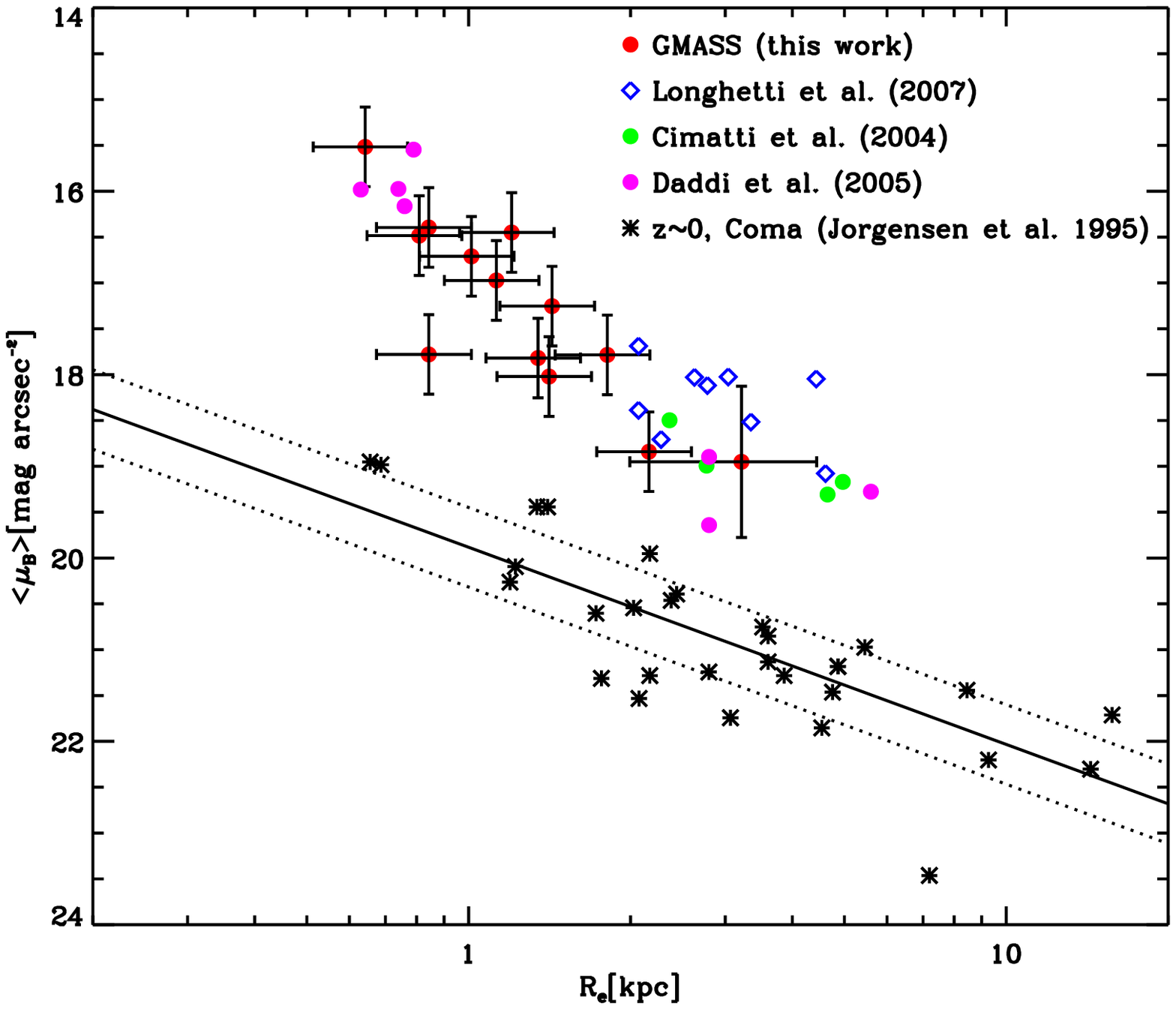}
\end{center}
\caption{ 
Mean surface brightness within $r_e$ in the B-band against $R_e$ for our
sample (red filled circles). Coma elliptical galaxies are shown as
stars, together with their best fit (solid line) and biweight scatter. Open
blue lozenges show $z\sim1.4$ galaxies from Longhetti~et~al.~(2007). 
}
\label{kormendy} 
\end{figure}

This relation for our sample is shown in Figure \ref{kormendy}. The
$B$-band rest-frame magnitude is measured during the photometric SED
fitting process described in Section 5.2. Here we applied the appropriate 
correction for the surface brightness dimming effect. The mean surface 
brightness $\mu_B$ is finally computed as:

\begin{equation}
\mu_B=M_B+DM(z)+2.5\log(2\pi)+5\log(r_e)+10\log(1+z),
\end{equation}

where $M_B$ is the absolute magnitude in the B-band, $DM(z)$ is the
distance modulus at redshift $z$, $r_e$ is the
effective radius in arcsec and $z$ is the redshift of the galaxy.

As a local ($z\sim0$) reference, we took the Kormendy relation for 
Coma galaxies by Jorgensen~et~al.~(1995). In Fig 18 we show both 
the individual Coma galaxies and the best fit to their relation
together with its $\pm 1 \sigma$ scatter. 
We also plot the $z\sim$1.4 galaxies of
Longhetti~et~al~(2007): their measurements (done on NICMOS images
and then converted into the rest-frame Gunn $r$ band) have been properly 
translated in B-band magnitudes applying a typical color 
$B-r$=1.4, and appear to be in good agreement with ours.
It can be noted that the Kormendy relation seems to evolve very strongly
between $z\sim$0 and $z\sim$1.5.
In particular, galaxies of the same size appear 
to be more luminous by an amount between 2 and 3 magnitudes. This brightening
is dependent on the physical size, being larger for smaller galaxies, thus
making the slope of the relation steeper at $z\sim 1.5$. In any case, all the
galaxies, if they have to end on top of the local relation, must fade their 
luminosity by more than 2 magnitudes from $z\sim$1.5 to $z\sim$0. This
fading exceeds by
at least 0.5 magnitudes the largest possible fading predicted by 
spectral evolution models of passive red galaxies (see also Longhetti 
et al. 2007 for a similar result). Since the surface mass density drops 
by a factor of 5-10 in the same redshift range, it means that these 
high-$z$ compact and dense galaxies can reach the local Kormendy relation 
only through a combination of increase in size ($R_e$) and fading due to
the passive evolution of the stellar populations.

\section{Are there superdense relics at $z\approx0$ ?}

The superdense ETGs identified at $z>1$ seem to disappear 
in the present-day Universe (e.g. Shen et al. 2003). However, 
recent works based on SDSS data 
showed that there is a population of very massive ETGs with 
$\sigma_V>350$ km s$^{-1}$ which, compared to ETGs of the same luminosity, 
are characterized by very small sizes (\cite{bernardi06,bernardi07}). 
Thus, we investigated whether some of these local galaxies have properties 
similar to those observed in passive galaxies at $z>1$. 

\begin{figure*}
\begin{center}
\includegraphics[width=\columnwidth]{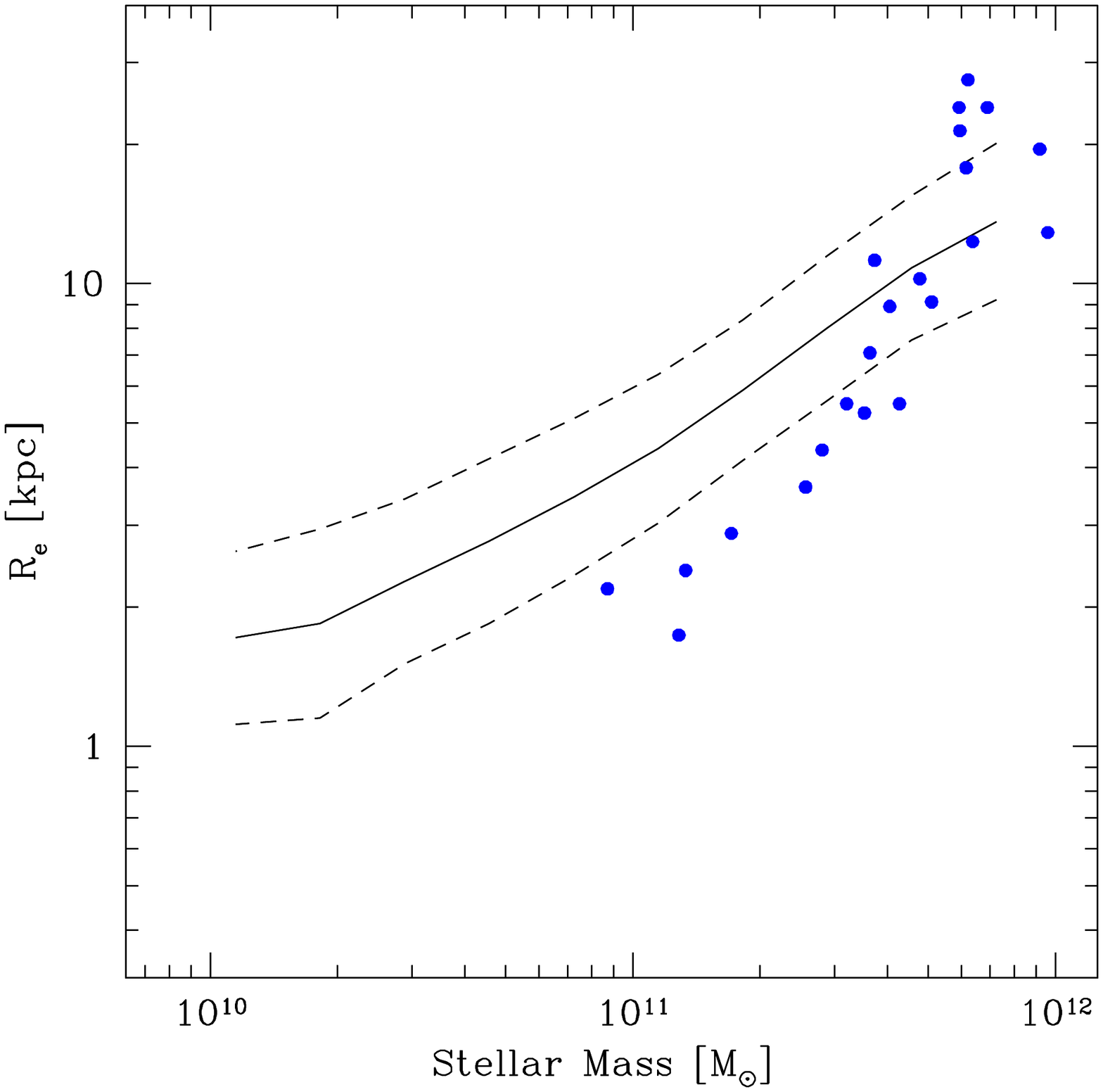}
\includegraphics[width=\columnwidth]{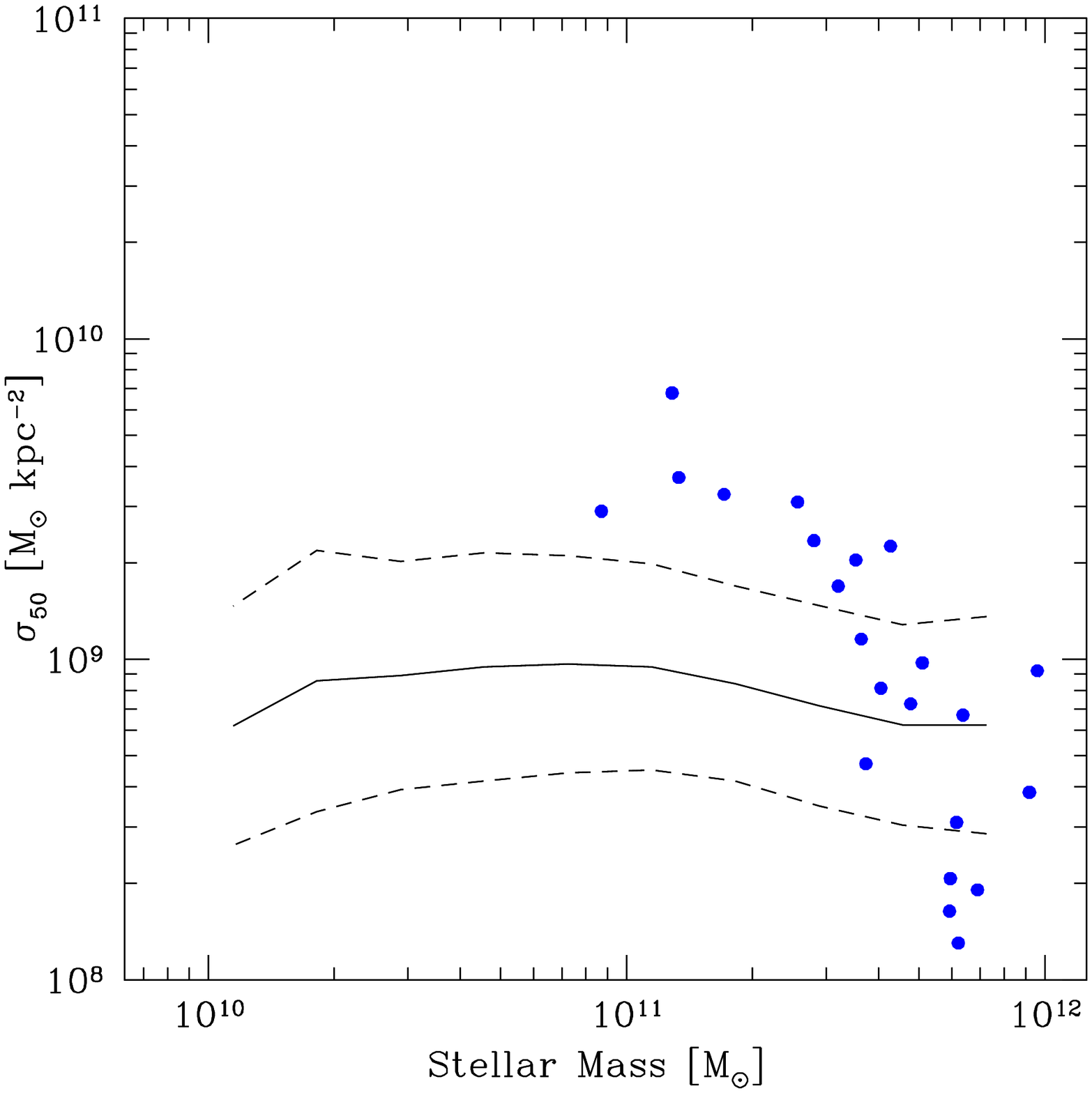}
\end{center}
\caption{The location of the ETGs at $z\approx0$ with $\sigma_V>350$ km
s$^{-1}$ identified by \cite{bernardi06,bernardi07} in the SDSS sample 
(blue circles) compared to the global relations of Shen et al. 2003 
(same as in Fig. 15--17).}
\label{mass-size} 
\end{figure*}

Fig. 19 shows the location of these galaxies in the $R_e$--mass and
$\sigma_{50}$--mass planes. The stellar masses derived with the method
described by Kauffmann et al. (2003) (that are also used by Shen et al. 2003)
were associated to the ETGs in the sample of Bernardi et al. (2007) (J. 
Brinchmann, private communication). Fig. 19 shows that there are a few
outliers which are not located within the Shen et al. (2003) relations and
are characterized by small sizes around 1--2 kpc with stellar masses
around $10^{11}$ M$_{\odot}$. These systems show a significant excess
in stellar mass surface density $\sigma_{50}$ (Fig. 19), but never
as strong as observed in ETGs at $z>1$. The
number density of these systems at $z\sim0$ is very small ($\approx
10^{-7}$--$10^{-8}$ Mpc$^{-3}$, \cite{bernardi06}), and much smaller
than the characteristic abundance of passive galaxies at $1<z<2$
($\approx 10^{-4}$ Mpc$^{-3}$, \cite{kong}). Thus, we conclude that
superdense relics with $R_e \approx$ 1 kpc at $z\approx0$ are extremely 
rare with respect to $z>1$, and absent if $R_e<1$ kpc.
However, in the "smooth envelope accretion" scenario discussed in 
Section 11.2.2,
it might be possible that the superdense galaxies observed at $z\approx1.5$ 
survived as hidden, very dense cores inside present-day ETGs.

\section{New constraints and open questions}

These findings provide new pieces in the puzzle of massive galaxy
formation and evolution, but at the same time open new crucial
questions: {\bf (1)} how did these systems form ? {\bf (2)} what
mechanism(s) can explain their subsequent size growth and decrease
of the internal density ?

\subsection{How did the superdense galaxies form ?}

The mere existence of $\approx$1 Gyr old, passively evolving galaxies 
up to $z\approx 2$ implies that the bulk of stars formed at higher redshifts. 
In particular, the constraints on the age, $\tau$ and stellar masses 
indicate that the star formation at $z>2-2.5$ was very intense (e.g. 
SFR$>100$ M$_{\odot}$ yr$^{-1}$) in order to form stellar mass up 
to ${\cal M} \gtrsim 10^{11}$ M$_{\odot}$ with such short timescales 
(see also \cite{cim04,mcc04,daddi05a,sar05,longhetti05,kriek}). 

The small sizes and high internal mass densities of the passive 
galaxies at $1<z<2$ provide new observational clues which help 
to unveil the nature of their precursors. Amongst all the precursor 
candidates mentioned in the Introduction, only the submm/mm-selected 
galaxies (SMGs) (\cite{blain02}) have \emph{sizes and mass surface density
comparable to those of passive galaxies} (\cite{tacconi,tacconi2}).

Table 5 compares the properties of SMGs and passive galaxies. 
The similarity between the properties of the two populations is 
striking, with the notable exception of the comoving number density,
with SMGs being an order of magnitude rarer (comoving number density 
$\approx 10^{-5}$ Mpc$^{-3}$; \cite{scott,chapman}) than the passive 
descendants ($\approx 10^{-4}$ Mpc$^{-3}$; \cite{daddi04,daddi05a,kong}).

The other populations 
of massive star-forming galaxies identified at $z \gtrsim $1.5--2 (e.g.
$BzK$, BM/BX, ...) show distinct properties if compared to SMGs: much 
larger sizes in the range of 2--8 kpc and mass surface densities 
lower by an order of magnitude 
($\approx 10^{2-3}$ M$_{\odot}$pc$^{-2}$) (see \cite{bouche}). 

The difference in these properties may reflect distinct dynamical
and assembly histories. SMGs may represent the cases where rapid and
highly dissipative major mergers occur at $z>2$ with timescales
of $\approx$0.1 Gyr (e.g. \cite{nara,ks06a,tacconi,tacconi2} and 
references therein) and leave very compact, superdense remnants
which then evolve almost passively at $1<z<2$ (see also \cite{swin}). 
The other star-forming systems selected in the optical and
near-infrared at $z>2$ may have a less violent evolution characterized
by star formation activity extended over longer timescales 
($\approx$0.5--1 Gyr, \cite{daddi05b}) and by multiple minor mergers or 
rapid dissipative collapse from the halo, with either process 
capable to form early disks. These massive disks may later also 
evolve into spheroids through disk instabilities of further merging 
processes (see \cite{genzel}). 

In this framework, objects like GMASS ID 2543 might represent the
transition phase of an object with a substantially
developed old stellar population concentrated in a bulge-like
structure which dominates the spectrum, SED and morphology, but
with still a residual star formation activity of a few solar 
masses per year as indicated by the weak [O II]$\lambda$3727 emission 
and the possible detection at 24$\mu$m. 

\begin{table}
\caption{Comparison between passive galaxies and SMGs}   % title of Table
\label{table:1}      % is used to refer this table in the text
\centering                          % used for centering table
\begin{tabular}{l | l | l}        % centered columns (4 columns)
\hline\hline                 % inserts double horizontal lines
& Quiescent galaxies & SMGs \\     
\hline                        % inserts single horizontal line
Redshift & $z \approx 1.5$ & $z \approx 2.5$ \\ 
Age of the Universe & 4.3 Gyr & 2.7 Gyr \\
Mass & $10^{10-11}$ M$_{\odot}$ (stars) & $10^{10-11}$ M$_{\odot}$ (gas) \\
Size & 1-2 kpc (stars) & 1-2 kpc (gas) \\

{\bf Mass surface density} & {\bf $10^{3-4}$ M$_{\odot}$pc$^{-2}$ (stars)} & {\bf $10^{3-4}$ M$_{\odot}$pc$^{-2}$ (total)}\\

Starburst timescale & 0.1-0.3 Gyr & $\approx$0.1 Gyr \\ 
Star formation rate & $<1$ M$_{\odot}$ yr$^{-1}$ & $\approx$100-1000 M$_{\odot}$ yr$^{-1}$ \\
Number density & $\approx 10^{-4}$ Mpc$^{-3}$ & $\approx 10^{-5}$ Mpc$^{-3}$ \\
Correlation length & $r_0 \approx 8-10$ Mpc & $r_0 \approx 6.9\pm 2.5$ Mpc \\
\hline                                   %inserts single line
\end{tabular}
\end{table}

Under the assumption that all passive galaxies at $z \approx 
1.5$ are the descendants of the starburst phase occurring in 
SMGs at $z \approx 2.5$, the characteristic timescale ("duty cycle" 
or the "duration") of the SMG phase can be estimated as the ratio 
of the comoving number densities of the two populations and the 
amount of cosmic time available from $z \approx 2.5$ to $z \approx1.5$ 
($\approx$1.5 Gyr), i.e. $\approx$0.15 Gyr. This timescale is broadly 
consistent with the $e$-folding timescale derived independently from 
the SED fitting and from the recent studies based on SMG molecular gas 
(\cite{tacconi,tacconi2}).

The evolutionary link between rapid, dissipational, gas--rich merging
or collapse (traced observationally by the SMG phase) at $z \approx$2--3 
and the 
superdense passive spheroids at $z \approx$1--2 provides one of the
strongest constraints known to date on the physical mechanisms 
capable to lead to the formation of massive spheroidal galaxies. 
Although the physical processes are completely different, this mechanism
is somehow reminiscent of the "old--fashioned" monolithic collapse. 

\subsection{How did the superdense galaxies disappear ?}

The other major question is to understand how the superdense galaxies
decreased their internal stellar mass density from $z\approx 1.5-2$ 
and migrated to the local size -- mass relation at $z\approx 0$.

\subsubsection{Dissipationless merging}

One possibility is the dissipationless ("dry") major merging of ETGs. 
According to some models (\cite{nipoti1,nipoti2,dominguez,naab07,
ciotti07,boylan}), this process can increase the final size and mass 
of the system without altering substantially the stellar population content. 
In particular, \cite{boylan} found that, under a set of orbital 
requirements, the stellar remnants of major dry mergers lie on 
the Fundamental Plane of their progenitors, and that the increase of 
the size with the increasing stellar mass is expected to follow a 
relation $R_e \propto {\cal M}^{\alpha}$, with $0.6 \lesssim \alpha \lesssim 
1.3$ depending on the orbital properties (see also \cite{nipoti2,ciotti07}).

Fig. 20 shows the expected growth of $R_e$ as a function of mass for the 
extreme values of $\alpha$, and suggests that $\alpha \gtrsim 1$ is needed
to grow efficiently the compact galaxies from $z\approx$1.5 and move
them onto the local size -- mass relation at $z\approx 0$. 
However, given all possible orbital parameters in dry merging events, we
may expect the {\it effective} value of $\alpha$ to be appreciably lower than
its maximum value. Therefore, the ability of dry merging alone to solve
the problem remains questionable.

\begin{figure}
\begin{center}
\includegraphics[width=\columnwidth]{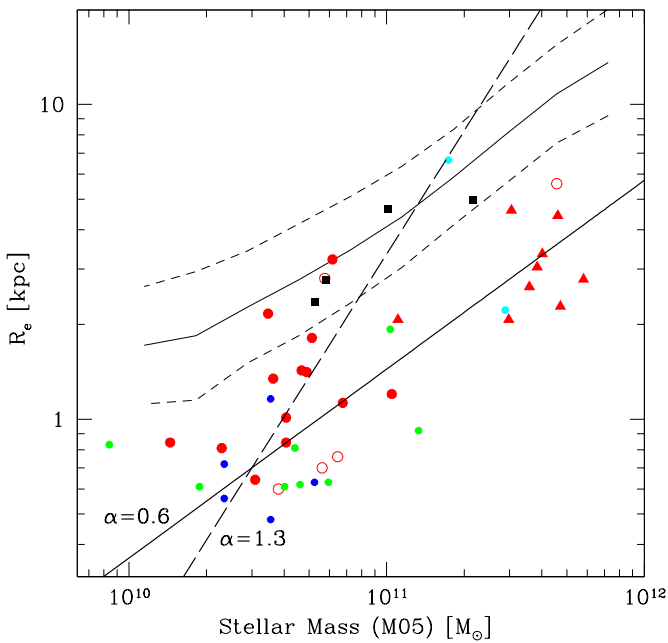}
\end{center}
\caption{The size--mass relation of passive galaxies at $z>1.3$.
The symbols are the same of Fig. 15. Stellar masses are estimated 
with M05 spectral synthesis models. The thin solid line shows the 
local size--mass relation of ETGs as derived with the SDSS sample 
by Shen~et~al.~2003, with the thin dashed lines indicating the 
$\pm 1 \sigma$ scatter around this relation. The thick solid and 
long--dashed lines indicate the expected increase of the size
due to dissipationless merging with $R_e \propto {\cal M}^{\alpha}$, 
with $\alpha=0.6$ and $\alpha=1.3$ respectively, and normalized to
$R_e$=0.7 kpc and a stellar mass of $3 \times 10^{10}$ M$_{\odot}$. 
}
\label{mass-size} 
\end{figure}

%Since $R_e$ of the superdense galaxies at $z\approx 1.5$ are on average
%a factor of $\approx$3 smaller than at $z\approx 0$, it means that
%typically each of these systems should experience $\approx$3 major 
%(i.e. $\approx$ equal mass) dry mergers (for $\alpha \approx$1) 
%in about 9 Gyr, i.e. with a rate of $\approx$0.3 Gyr$^{-1}$.
The dry merging hypothesis can in principle be tested by 
deriving the statistics of pairs of close/interacting passive 
galaxies as a function of redshift. The redshift range $1<z<2$ 
seems to be the most important one because of the substantial decrease of 
the superdense ETG fraction from $z\approx$1.4--2 to $z\approx$0.6--1.3 
(Fig. 17 vs Fig. 15; see also \cite{mci,trujillo07}). Unfortunately, no dry 
merger statistics is available in this redshift range (see e.g. 
\cite{bell06}). We also recall that to be fully valid, the dry merging scenario 
should also be consistent with the recent findings of a very weak
evolution of the stellar or dynamical mass functions of luminous and
massive ($>10^{11}$ M$_{\odot}$) ETGs at $z \leq$0.7--0.8 (e.g. 
\cite{bundy07,bundy06,borch,cdr06,scarlata,pozz07}).

Due to the limited statistics of the present GMASS sample, it is 
not possible to place 
any strong constraint on the dry merging scenario. However, we note that 
12 out of 13 galaxies (see Fig 11) have no visible companions within a 
distance of about 10-15 kpc. The only, exception is represented by ID 996 
which has a companion galaxy located at a small distance,
but neither spectroscopic nor photometric redshift are available for 
this galaxy. The only available constraint comes from the color 
($F110W-F160W \approx 0.6$) which looks similar to that of ID 996 
($F110W-F160W \approx 0.8$). However, even if this galaxy is at the 
same redshift of ID 996 and it will merge with it, this would not 
represent a "major" merger as the companion is fainter than 
ID 996 by $\approx 2.5$ magnitudes, corresponding to a luminosity 
(or $\approx$mass) ratio of $\approx$1:10. 

Even if clear cases of dry merger candidates are not present in our 
sample, we note that about half of the passive galaxies found in 
the GMASS spectroscopic sample lie in the dense structure at $z=1.61$ 
(Vanzella et al. 2006; Castellano et al. 2007; Kurk et al. 2007b). 
The 7 galaxies present in this redshift spike
have typical angular separations among themselves of $\approx$ 10 arcsec, 
corresponding to $\approx$ 85 kpc. It is tempting to speculate that 
some of these galaxies will later merge with each other along the filaments
of this structure and form one or more larger massive ETG 
at lower redshift (see e.g. Nipoti et al. 2003b). McCarthy et 
al. (2007) have recently identified a compact cluster or group of
red (passive) galaxies at $z=1.5$ which may represent 
an example of a short--lived phase leading to the subsequent assembly
of more massive ETGs at later cosmic times.

In this framework, further constraints can come from the differential
evolution of the size--mass relation of ETGs as a
function of the environment. We preliminarily investigated this issue
using the sample of Rettura et al. (2006) which includes literature 
cluster and field
ETGs at $z\approx$0.6--1.3. The sample is small and by no means complete.
However, Fig. 21 suggests that environmental effects might be an
important component of the size--mass relation evolution. Cluster ETGs
seem to be preferentially located within or closer to the size--mass
relation at $z\approx0$ with respect to the ETGs located in low density
environment. If confirmed with larger and complete samples, this
may be consistent with the scenario where the mass assembly and size 
growth of ETGs is accelerated within high--density environment.

\begin{figure}
\begin{center}
\includegraphics[width=\columnwidth]{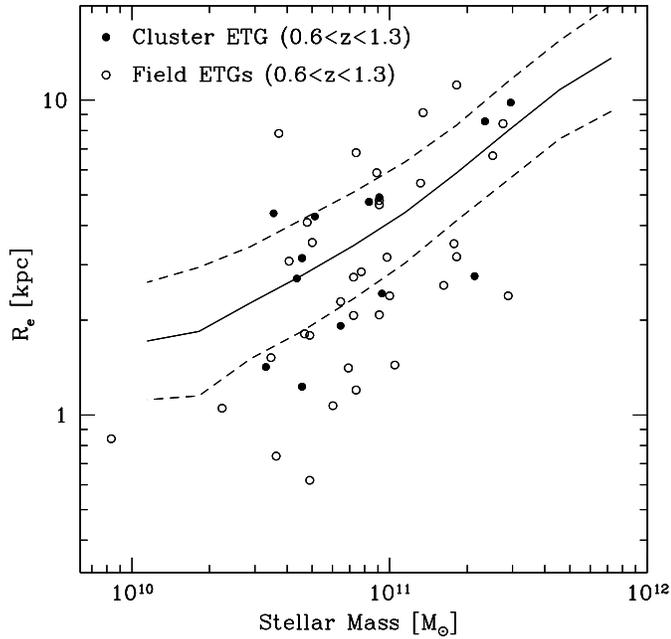}
\end{center}
\caption{The distribution of physical sizes versus stellar mass for
ETGs at $z\approx1$ of Rettura et al. (2006). 
The solid line shows the local size--mass relation of ETGs 
(Shen~et~al.~2003), with the dashed lines indicating the 
scatter around this relation. The filled and open symbols indicate cluster
and field ETGs respectively. Stellar masses are
homogeneously estimated with the PEGASE model spectra (Kroupa IMF).}
\end{figure}

Recent simulations attempted to explain the size evolution of ETGs
in the scenario of dissipationless merging.
Khochfar \& Silk (2006b) used semianalytical models to show 
that the size of elliptical galaxies (relative to $z \approx 0$) depends
on the available amount of cold gas and the fraction of stars formed 
during the major merger event with which these galaxies formed. In 
this scenario, massive galaxies formed at high redshifts through 
gas-rich mergers which produced elliptical "remnants" characterized 
by small sizes. The size evolution is stronger for more massive galaxies 
as they involve more gas at high redshifts when they form, compared 
to less massive ellipticals, and galaxies of the same mass at low 
redshift form mostly from gas-poor mergers (see also Khochfar \& Silk 2006a). 
According to this picture, local ellipticals with {\it present--day}
stellar masses $10^{10}< {\cal M}(z=0) < 10^{11}$ M$_{\odot}$ were only 
$\approx$1.25 times smaller at $z=2$, whereas those with ${\cal M}(z=0)> 
5 \times 10^{11}$ M$_{\odot}$ were $\approx$4 times smaller at $z=2$. 
Khochfar \& Silk (2006b) also predict that the most massive ellipticals 
undergo on average 1--2 substantial "dry" mergers between $z=2$ and today.
%not so different from the rough estimate of $\approx$3 discussed above.

Figure 22 shows the location of the GMASS passive galaxies with
respect to the model predictions of Khochfar \& Silk (2006b). The ratio
$R_e(z) / R_e(z=0)$ was obtained dividing the observed effective radius
of the GMASS galaxies by that at $z=0$ derived from the stellar 
mass -- size relation of SDSS ETGs of Shen et al. (2003)
at the same mass of the GMASS passive galaxies. M05 stellar masses
are used in this comparison, but the results do not change
significantly if BC03 or CB07 masses are used. According to this
comparison, the model predictions suggest that most GMASS passive
galaxies are the progenitors (or the high-$z$ counterparts) of
ETGs that \emph{today} have stellar masses of $10^{11}< {\cal M}(z=0) < 
10^{12}$ M$_{\odot}$, i.e. they are the progenitors
of the most massive E/S0 systems at $z\approx0$.

\begin{figure}
\begin{center}
\includegraphics[width=\columnwidth]{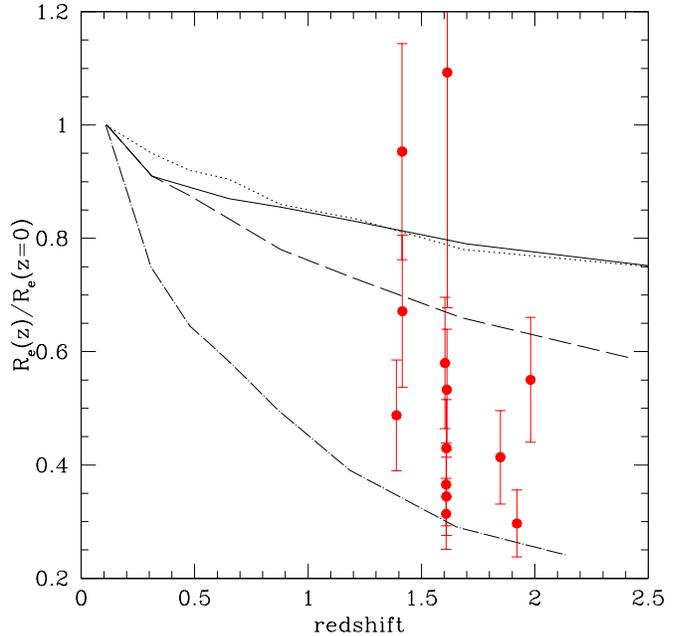}
\end{center}
\caption{The ratio of $R_e$ of GMASS passive galaxies to
the mean size of $z\approx0$ ETGs, 
$\langle R_e(z,{\cal M})/R_e(0,{\cal M}) \rangle$, as derived
from Shen et al. (2003). M05 stellar masses are used here.
The curves represent the predictions of the
Khochfar \& Silk (2006b) model relative to the {\it final masses at $z=0$}.
Solid, dotted, dashed, dotted - dashed lines are relative respectively 
to the following ranges of stellar masses
at $z\approx0$: $1 \times 10^{10}< {\cal M}(z=0) < 5 \times 10^{10}$ M$_{\odot}$,
$5 \times 10^{10}< {\cal M}(z=0) < 1 \times 10^{11}$ M$_{\odot}$,
$1 \times 10^{11}< {\cal M}(z=0) < 5 \times 10^{11}$ M$_{\odot}$,
$5 \times 10^{11}< {\cal M}(z=0) < 1 \times 10^{12}$ M$_{\odot}$.
}
\end{figure}

\subsubsection{Smooth envelope accretion}

With an alternative approach based on smoothed particle hydrodynamics
(SPH) simulations, Naab et al. (2007) showed that ETGs can
be formed with appropriate cosmological initial conditions and even
without requiring recent major merger events or feedback from supernovae
or AGN. These models also predict small sizes around 1-2 kpc at $z>1$. 
The Naab et al.
(2007) approach is also relevant because it shows that during the early
formation at $z>2$ the assembly of massive ellipticals is dominated by
mergers of gas-rich subcomponents and in situ star formation in a way
characteristic of a dissipative collapse. Thereafter, smooth stellar accretion
or minor/major mergers become more important and tend to dominate
at $z<1$, although major mergers are not always necessarily needed. 
In these simulations, the size evolution is mostly driven by the
{\it accreted stars}. While the half-mass radius of the stars formed in situ
(at high redshifts) remains almost constant independent of redshift, 
the accreted stars form an {\it envelope} whose half-mass radius 
increases smoothly with decreasing redshift around the compact and dense
"seed" formed at higher redshifts. If this scenario is correct, we may 
expect to find the compact, superdense cores "hidden" inside 
present-day ETGs. 

\section{Summary and outlook}

We presented a study of a 4.5$\mu$m--selected sample of passive galaxies 
spectroscopically identified at $z\approx$1.4--2 with the GMASS survey 
(Kurk et al. 2007a). This work benefits from ultradeep VLT+FORS2 optical 
spectroscopy complemented with multi-band photometry (0.4--8$\mu$m)
and HST imaging to investigate the physical, structural and evolutionary 
properties of these galaxies. The main results can be summarized as follows.

$\bullet$ The GMASS passive galaxies have spectra and SEDs dominated
by old stars and very weak or absent star formation. A comparison of the
stacked rest-frame UV spectrum (equivalent to an integration time of 
$\sim$500 hours) with three different libraries of stellar population 
model spectra indicates an age of $\approx$0.7--2.8 Gyr for a
metallicity range of 1.5--0.2 $Z_{\odot}$. Extending the model fitting 
at longer wavelengths using near-infrared and IRAC photometry
helps to reduce the age--metallicity degeneracy and 
indicates ages of $\approx$1--1.6 Gyr, $Z=Z_{\odot}$, $e$-folding 
timescales $\tau \sim$0.1--0.3 Gyr, where $SFR(t) \propto exp(-t/ \tau)$,
and very low dust extinction. A comparison of the two libraries of model 
spectra which include TP-AGB stars shows that for a fixed age of 1 Gyr, 
the Maraston 
(2005) templates show a better agreement with the photometry than those 
of Charlot \& Bruzual (2007) available to date.

$\bullet$ Neither individual galaxies nor their stacked images are detected 
in the X-rays with \textit{Chandra} data. This implies that a luminous 
AGN source ($L_X>10^{42}$ erg/s) is absent or is very heavily obscured. 
However, the possibility of dust obscuration seems unlikely because none 
of the galaxies has been detected at 24$\mu$m, with only one marginal
exception of one galaxy at $z=1.61$. 

$\bullet$ The stellar masses, estimated through the photometric SED 
fitting, are in the range of $10^{10-11}$ M$_{\odot}$ and the specific
star formation rates are very low ($\lesssim 3 \times 10^{-2}$ Gyr$^{-1}$). 
The stellar masses estimated with model spectra including TP-AGB stars 
are systematically lower by 0.1--0.2 dex than those estimated with 
models which do not include this phase of stellar evolution.

$\bullet$ The HST+ACS morphological and surface brightness profile analysis 
indicate that the majority of the spectroscopically--selected passive 
galaxies have spheroidal morphologies consistent with being analogous
to present-day ETGs. However, their sizes are smaller by a factor 
of $\approx$2-3 than at $z\approx0$, and imply that the stellar mass surface 
and volume internal densities are up to $\approx$10 and $\approx$ 30 times 
larger respectively. If literature data are added to the GMASS sample,
we find that only a few passive systems at $1.2<z<2.5$ lie 
within the $z\approx0$ size -- mass relation, whereas the majority 
has systematically much smaller sizes. 

$\bullet$ The Kormendy relation at $z\approx1.5$ shows a large offset 
(2--3 mag) in effective surface brightness with respect to the local 
relation. This is difficult to explain with simple luminosity evolution 
models and requires that these high-$z$ compact and dense galaxies 
increase their size in order to reach the local Kormendy relation.

$\bullet$ Samples of ETGs at lower redshifts ($0.7<z<1.2$) show that a 
larger fraction of passive galaxies follow the $z\sim0$ size -- mass 
relation with respect to $z>1.3$. The ETGs at $z\approx1$ which have the
largest offsets with respect to the $z\sim0$ size -- mass relation are
the ones having the highest internal velocity dispersion, as expected
from the ETG scaling relations. We find a hint that, for a fixed
redshift $z\approx$1, ETGs located within massive clusters are
more preferentially located within the $z\sim0$ size -- mass relation
than ETGs located in lower density environments.

$\bullet$ Superdense massive ETGs with $R_e \approx$ 1 kpc are extremely rare 
at $z\approx0$ with respect to $z>1$, and absent if $R_e<1$ kpc. However,
it might be possible that compact and dense remnants are "hidden" inside
present-day ETGs if the size of ETGs grew through mechanisms such as
the "smooth envelope accretion".

$\bullet$ Submillimeter--selected galaxies are the only systems at 
$z\gtrsim$2 with sizes and mass surface densities (in gas) similar to those
of the passive galaxies at $z\approx$1--2. This suggests that a strong
evolutionary link is present between these two galaxy populations.

$\bullet$ It is currently unclear how the possible link between SMGs
and compact passive galaxies fits within a more general framework which 
takes into account also the other galaxy populations so far identified at 
$1<z<3$. A plausible scenario could be outlined as follows and summarized
in Fig. 23. {\bf (1)} Massive star-forming galaxies 
selected in the optical/near-IR are gas--rich disky systems with 
long--lived star formation (e.g. $\approx$0.5--1 Gyr, \cite{daddi05b}). 
{\bf (2)} These systems can become unstable (e.g. \cite{genzel}) or 
participate in major merger events with other gas-rich systems. 
{\bf (3)} In both cases a major starburst event is triggered, and this 
phase could correspond to the SMG stage characterized by short-lived 
($\approx$0.1 Gyr) vigorous starburst (\cite{tacconi,tacconi2}). {\bf 
(4)} The concomitant AGN provides enough feedback to ``quench'' the 
star formation in massive systems (Daddi et al. 2007a), and {\bf (5)} 
compact, superdense, passively evolving remnants are formed, {\bf (6)} and
evolve subsequently by increasing gradually their sizes with mechanisms 
like major dry merging and/or envelope accretion more or less rapidly 
depending on their mass and environment. {\bf (7)} The 
majority of most massive ETGs reach the assembly completion around 
$z\approx$0.7, while lower mass ETGs continue to assemble down to
lower redshifts (downsizing). 

\begin{figure}
\begin{center}
\includegraphics[width=\columnwidth]{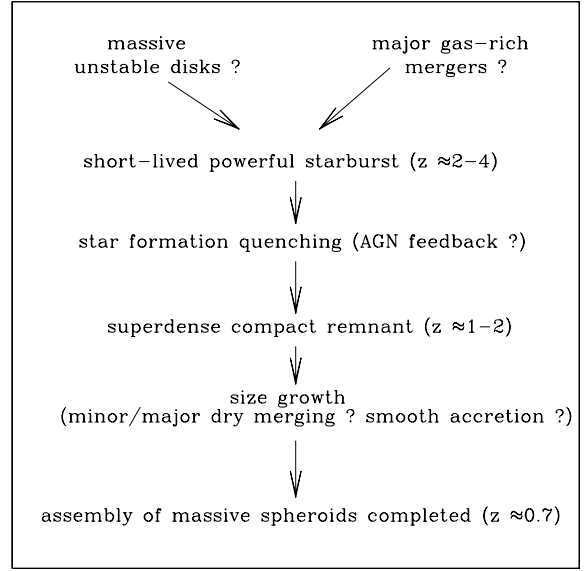}
\end{center}
\caption{A possible scenario for the formation and evolution of a
massive spheroidal galaxy (see text).}
\end{figure}

$\bullet$ Larger samples in the range of $1<z<3$ are needed to place
stringent constraints on the mechanism(s) with which the sizes of
high--$z$ passive galaxies grow as a function of cosmic time (e.g.
dissipationless merging, envelope accretion, ...). High--resolution
spectroscopy in the near--infrared with the next generation of
telescopes (e.g. E-ELT, JWST) will also be crucial to study
the internal dynamics of these systems. Further studies at $1<z<3$ 
will shed light on the global processes which lead to formation and 
evolution of massive galaxies.

\begin{acknowledgements}
We acknowledge the referee, Olivier Le Fevre, for the useful and
constructive comments.
We thank ESO for the generous allocation of observing time through
the VLT Large Program 173.A--0687. We thank Mariangela Bernardi, 
Joel Brinchmann, Pat McCarthy, Georg Feulner, and Sadegh Khochfar for 
providing respectively unpublished information on SDSS galaxies with
$\sigma_V>$350 km s$^{-1}$, the SDSS stellar masses,
the GDDS composite spectra, the data on SSFR, and the model predictions. 
Sadegh Khochfar, Thorsten Naab, Luca Ciotti, Carlo Nipoti and Gabriella
De Lucia are acknowledged for the helpful discussion. AC is particularly
grateful to Reinhard Genzel, Linda Tacconi and Natascha F\"orster-Schreiber for 
the warm hospitality at Max-Planck-Institut f\"ur Extraterrestrische Physik 
and the stimulating discussions. AC acknowledges support through a 
\emph{Bessel Prize} of the \emph{Alexander von Humboldt Foundation}.
\end{acknowledgements}

\end{document}